%% file: main.tex
\pdfoutput=1
\documentclass[10pt,conference]{IEEEtran}
\usepackage{cite}
\usepackage{amsmath,amssymb,amsfonts}
\usepackage{algorithmic}
\usepackage{graphicx}
\usepackage{textcomp}
\usepackage{xcolor}
\usepackage{array}
\usepackage{booktabs}
\newcommand{\tablecolsep}{{\color{black!30}\vrule width 0.25pt}}
\usepackage{tikz}
\usetikzlibrary{patterns,arrows.meta} %
\usepackage{placeins}
\usepackage{stfloats}
\usepackage{balance}
\usepackage[hyphens]{url}
\usepackage{fancyhdr}
\usepackage[hidelinks]{hyperref}
\newcommand{\figpanel}[2]{\hyperref[#1]{Figure~\ref*{#1}#2}}

\renewcommand{\footnoterule}{%
  \kern-3pt
  \hrule width 0.25\columnwidth height 0.4pt
  \kern2.6pt}

\title{Should You Split Decode? Disaggregation
Across Future Models and Accelerators}
\newcommand{\tool}{\textsc{SiliconForge}} %

\IEEEoverridecommandlockouts
\IEEEaftertitletext{\vspace{-1pt}}
\author{\IEEEauthorblockN{Ramyad Hadidi$^{\dagger}$\thanks{$^{\dagger}$Ramyad Hadidi wrote this paper independently, in personal time while at d-Matrix and at Volantis Semiconductor and in the interval between the two. The tools, code, and parameters contain no confidential information. All views are the author's own and do not represent those of any employer.}}}

\begin{document}
\maketitle
\thispagestyle{plain}
\pagestyle{plain}

\input{sections/0-Abstract}
\input{sections/1-Intro}
\input{sections/2-Tool}

\input{sections/3-Inputs}
\input{sections/4a-Insights}

\input{sections/4b-Insights}
\input{sections/5-Calibration}
\input{sections/6-Discussion}
\input{sections/8-RelatedWork}
\input{sections/9-Conclusion}

\clearpage
\IEEEtriggercmd{\balance}
\IEEEtriggeratref{144} %
\bibliographystyle{IEEEtranS} %
\bibliography{refs} %

\input{sections/13-AIUse}
\end{document}

%% file: sections/0-Abstract.tex
\begin{abstract}
For large language model (LLM) inference, prefill/decode disaggregation (P/D) is
deployed in production to isolate phases with different resource demands.
Attention/feed-forward network disaggregation (AFD) splits decode again, but
adds communication, placement constraints, and another provisioned pool. We
ask whether role-specific placement repays this cost as models and accelerator
roadmaps change. Infrastructure is committed before future models and hardware
can be benchmarked.
We present \tool{}, a calibrated model that varies the serving decomposition
and hardware assignment independently. We compare P/D and P/D+AFD across four
model directions, four accelerator classes anchored to current and announced
parts, and two fabrics. We co-optimize each design using the same rules for
pipeline overlap and the same power budget.
Our main comparisons show no robust deployment-level AFD advantage
in service-level objective (SLO) goodput per kW. Most nominal
estimates favor P/D. The largest AFD lead
comes from fitting one more prefill replica, not from a more efficient decode
pool. AFD has a nominal lead in the KV-heavy (key--value) model direction when the co-located
expert pool spans several small scale-up domains. Larger domains or a fast fabric for
both designs weaken that lead.
We therefore find that prefill provisioning, decode memory bandwidth, links,
and pool granularity decide deployments more than the split alone.
A fixed hardware assignment can lose its advantage as models, fabrics, hardware
assumptions, and objectives change.
Re-optimizing mappings and pool sizes on the same hardware cuts early commitment's median normalized goodput loss from 36\% to 9\%. We recommend adjustable mappings and pool sizes to let boundary and hardware choices follow model demands.
\end{abstract}

%% file: sections/1-Intro.tex
\section{Introduction}\label{sec:intro}

Accelerator roadmaps for large language model (LLM) inference trade compute, memory bandwidth, capacity, interconnect, and power, while prefill, decode attention, and decode feed-forward computation stress those resources differently. Prefill is compute-intensive and sets time to first token (TTFT). Decode attention reads a growing key--value (KV) cache, and the feed-forward network (FFN) reads dense or expert weights, together setting time per output token (TPOT)~\cite{zhong2024distserve,patel2024splitwise,qin2025mooncake,zhu2025megascaleinfer}. This mismatch creates two decisions. The \emph{serving decomposition} chooses which roles get their own \emph{pool} of accelerators. The \emph{hardware assignment} selects an accelerator class for each pool. A \emph{boundary} is where data crosses the network between pools. Prefill/decode disaggregation (P/D) separates prefill from decode; attention/feed-forward network disaggregation (AFD) splits decode attention from the FFN.

P/D is standard industry practice~\cite{chen2025distserveretro}. It is supported by serving frameworks such as NVIDIA Dynamo, vLLM, and SGLang~\cite{elmeleegy2026dynamo1,vllm2026disaggprefilldocs,sglang2026pddocs}. AFD is also being seriously considered. ByteDance reports AFD in production on nearly 10,000 GPUs~\cite{zhu2025megascaleinfer}. NVIDIA's Vera Rubin platform assigns decode attention to Rubin GPUs and mixture-of-experts (MoE) FFNs to Groq~3 LPX accelerators~\cite{nvidia2026groq3lpx}.

AFD adds a new serving boundary. Each boundary moves data, creates independently sized pools, can strand capacity, and restricts placement. AFD helps only if role-specific placement repays those costs. However, AFD sends each token across the boundary twice per MoE layer, or 116 times in DeepSeek-V3~\cite{deepseek2024v3}. In contrast, co-located decode with expert parallelism keeps expert communication within the decode pool, but the KV cache and expert weights share device memory. This can limit batch size as contexts grow~\cite{zhu2025megascaleinfer}. Prior studies often change the boundary and hardware assignment together. Studying the two choices independently separates the effect of splitting decode from the effect of hardware assignment. Whether or not AFD becomes the preferred design, these decisions commit substantial capital, determine latency and throughput, and influence accelerator roadmaps. For accelerators still on the roadmap, hardware design changes become expensive once the silicon design is frozen~\cite{synopsys2026platformarchitect}. We ask one question. \emph{Does AFD repay its cost across future models and accelerators, and which lever matters more?}

To answer this question, we compare P/D and P/D+AFD and their hardware assignments before future models and accelerators are built. Existing AFD measurements come from system builders or hardware vendors, each using their own hardware and baseline~\cite{zhu2025megascaleinfer,fu2026fastafd,nvidia2026groq3lpx}. However, a prospective comparison must account for future model demands, workload ranges, and changes in accelerator capabilities. We built \tool{}, a calibrated cross-layer model that composes LLM architectures with hardware specifications. To build \tool{}, we assemble a catalog of 39 accelerators in six hardware archetypes and define four classes for the comparison (Table~\ref{tab:hardware}). Drawing on six published DeepSeek, Kimi, and Qwen models, we construct four prospective directions by varying expert and KV demands while rescaling width or depth (Table~\ref{tab:models}). \tool{} models compute and memory costs, communication, and pipeline overlap (\S\ref{sec:tool-demand}--\S\ref{sec:tool-metrics}, \S\ref{sec:model-equations}). Queueing converts the resulting service times into goodput, while an event-driven simulation checks the queueing model. We calibrate against published measurements and check comparative accuracy on published comparison pairs (\S\ref{sec:validation}). \tool{} reports several metrics, including saturation throughput and service-level objective (SLO) goodput, the largest request arrival rate that meets both TTFT and TPOT targets~\cite{zhong2024distserve}.

Both designs share compute, memory, pipeline, and queueing models and a calibration policy fixed across the design space. Updating the models and calibration as future LLM and hardware specifications become clearer lets the community evaluate new designs. We independently optimize each design's mapping and pool sizes under a common power budget. A margin passes our two-part test only if its nominal magnitude exceeds the 9\% envelope and its sign holds at every tested specification corner.

Our first finding is that \textbf{(I)} AFD has no robust deployment-level advantage in SLO goodput per kW across the accelerator classes and model directions in our main comparison. Most nominal estimates favor P/D. The largest AFD lead comes from fitting one more prefill replica into the budget. When decode alone is optimized, P/D+AFD has lower throughput per decode kW (\S\ref{sec:regime}). Our second finding is that \textbf{(II)} AFD has a nominal lead in the KV-heavy model direction when the co-located expert pool spans several small scale-up domains. Larger domains or a fast fabric for both designs weaken that lead. More broadly, prefill provisioning, decode memory bandwidth, links, and pool granularity decide deployments more than the split alone (\S\ref{sec:capacity}). With decode mappings fixed, doubling memory bandwidth cuts TPOT by 39--49\%, while doubling FLOPs cuts it by at most 1.8\%. Prefill traffic also makes link bandwidth matter beyond the attention/FFN exchange (\S\ref{sec:regime}, \S\ref{sec:rolehw}). Our third finding is that \textbf{(III)} a fixed hardware assignment can lose its advantage as models, fabrics, hardware assumptions, and objectives change. Re-optimizing mappings and pool sizes on the same hardware cuts early commitment's median loss of normalized goodput from 36\% to 9\%. The tested reserves add less recovery (\S\ref{sec:regret}).

We make three contributions. First, \textbf{\tool{}, a cross-layer model}, compares independently optimized P/D and P/D+AFD under common modeling assumptions and a common power budget. We check comparative accuracy on held-out published pairs for each boundary and report failures (\S\ref{sec:validation}). Second, we identify \textbf{when splitting decode helps}, showing how prefill provisioning, decode memory bandwidth, links, scale-up domains, and pool granularity govern its modeled benefit (\S\ref{sec:regime}, \S\ref{sec:rolehw}). Third, we quantify \textbf{the goodput cost of early hardware commitment} and the recovery available from re-optimizing mappings and pool sizes (\S\ref{sec:regret}).

%% file: sections/2-Tool.tex
\section{Disaggregated Serving for LLMs}\label{sec:tool}\label{sec:background}

Disaggregated LLM serving assigns request work to separate pools of accelerator replicas with independently chosen hardware, size, and batching~\cite{zhong2024distserve,patel2024splitwise,qin2025mooncake}. P/D places prefill and decode in separate pools. P/D+AFD further separates decode attention and FFN into distinct pools. This section explains \tool{}'s modeling and metrics.

\begingroup
\setlength{\parskip}{0pt}
\renewcommand{\subsection}{\csname @startsection\endcsname{subsection}{2}{0pt}{1.5ex}{0.7ex}{\normalfont\normalsize\itshape}}
\subsection{Serving pools and boundaries}
\label{sec:tool-knobs}

A serving boundary is a handoff between pools, determining which work shares hardware and which crosses the fabric. Unified serving keeps prefill and decode in one pool. Production systems limit interference with chunked prefill and decode priority~\cite{agrawal2024sarathiserve}. P/D places the two phases in separate pools and transfers the completed prompt KV state once per request (the \emph{KV handoff})~\cite{zhong2024distserve,patel2024splitwise}. A request's KV cache holds its \emph{resident context}. The \emph{capacity-bound batch} is the maximum number of requests whose weights and caches fit in usable memory and therefore shrinks as context grows.
\par
\endgroup

In an MoE layer, each token activates $k$ of $E$ experts~\cite{shazeer2017moe} (8 of 256 routed experts in DeepSeek-V3). Expert parallelism (EP) spreads experts over decode units, sending each token to its experts and back (in-pool dispatch and combine). P/D co-locates attention and FFN/expert work within the decode pool, sharing devices and a batch. P/D+AFD separates an attention pool that holds and reads KV state from a dense or expert FFN pool. Activations cross the \emph{inter-pool link} twice per MoE layer of every decode step, once each way (the \emph{exchange})~\cite{zhu2025megascaleinfer,wu2026howfar}. Attention reads KV state, while experts repeatedly stream weights, creating different capacity and bandwidth demands. Role-specific placement can match them but adds recurring communication and idle capacity.

\begingroup
\setlength{\textfloatsep}{2pt}
\setlength{\parskip}{0pt}
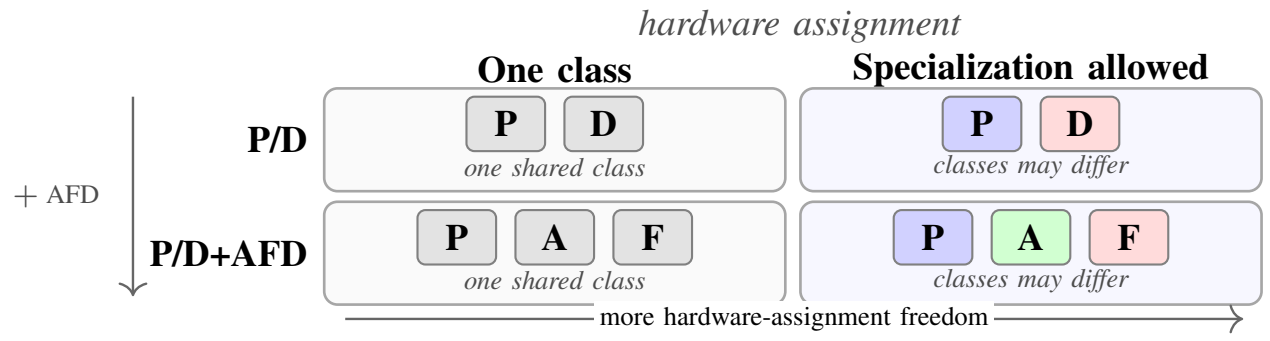
\begin{figure}[!b]
\centering
\makebox[\linewidth][r]{%
\begin{tikzpicture}[xscale=0.9,yscale=0.75,
  config/.style={draw=black!35, rounded corners=2.5pt, line width=0.5pt,
    minimum width=3.05cm, minimum height=0.68cm, inner sep=2pt},
  role/.style={draw=black!50, rounded corners=1.5pt, line width=0.4pt,
    minimum width=0.52cm, minimum height=0.36cm, inner sep=1pt,
    font=\scriptsize\bfseries},
  cell note/.style={font=\tiny\itshape, text=black!70},
  column label/.style={font=\scriptsize\bfseries, align=center},
  row label/.style={font=\scriptsize\bfseries, anchor=east},
  axis arrow/.style={->, draw=black!60, line width=0.45pt}
]
  \node[font=\scriptsize\itshape, text=black!70] at (4.50,2.48)
    {hardware assignment};
  \node[column label] at (2.70,2.10) {One class};
  \node[column label] at (6.20,2.10) {Specialization allowed};

  \node[row label] at (1.02,1.48) {P/D};
  \node[row label] at (1.02,0.48) {P/D+AFD};

  \node[config, fill=gray!4] at (2.70,1.48) {};
  \node[role, fill=gray!22] at (2.34,1.62) {P};
  \node[role, fill=gray!22] at (3.06,1.62) {D};
  \node[cell note] at (2.70,1.24) {one shared class};

  \node[config, fill=blue!3] at (6.20,1.48) {};
  \node[role, fill=blue!18] at (5.84,1.62) {P};
  \node[role, fill=red!14] at (6.56,1.62) {D};
  \node[cell note] at (6.20,1.24) {classes may differ};

  \node[config, fill=gray!4] at (2.70,0.48) {};
  \node[role, fill=gray!22] at (1.98,0.62) {P};
  \node[role, fill=gray!22] at (2.70,0.62) {A};
  \node[role, fill=gray!22] at (3.42,0.62) {F};
  \node[cell note] at (2.70,0.24) {one shared class};

  \node[config, fill=blue!3] at (6.20,0.48) {};
  \node[role, fill=blue!18] at (5.48,0.62) {P};
  \node[role, fill=green!18] at (6.20,0.62) {A};
  \node[role, fill=red!14] at (6.92,0.62) {F};
  \node[cell note] at (6.20,0.24) {classes may differ};

  \draw[axis arrow] (-0.40,1.86) -- (-0.40,0.10);
  \node[font=\tiny, anchor=east, text=black!70] at (-0.51,0.98)
    {$+\,$AFD};
  \draw[axis arrow] (1.16,-0.10) -- (7.76,-0.10);
  \node[font=\tiny, fill=white, inner sep=1pt]
    at (4.46,-0.10) {more hardware-assignment freedom};
\end{tikzpicture}%
}
\par\vspace{-9pt}
\caption{\textbf{Serving decomposition and hardware assignment.} P, D, A,
and F denote prefill, decode, attention, and FFN.}
\label{fig:inputs-design-axes}
\end{figure}

Figure~\ref{fig:inputs-design-axes} shows serving decomposition and hardware assignment as independent choices. Adding a boundary changes communication, pooling, capacity, and queueing even when the two pools it separates use the same accelerator class. Reassigning hardware changes service time, feasibility, and provisioned power without changing the boundary. We therefore vary the two axes separately. For each experiment, we fix the workload, precision, fabric, SLO, and queueing convention. A mapping sets tensor parallelism (TP), data parallelism
(DP), EP, pipeline parallelism (PP), the per-replica batch,
and the microbatch count $M$. For P/D and P/D+AFD separately,
\tool{} searches the allowed hardware assignments, mappings,
and pool sizes (\S\ref{sec:inputs}). It maximizes SLO goodput across feasible deployments under the common power budget.

\begingroup
\clubpenalties 2 10000 0
\renewcommand{\subsection}{\csname @startsection\endcsname{subsection}{2}{0pt}{1.5ex}{0.7ex}{\normalfont\normalsize\itshape}}
\subsection{LLM model demands on serving pools}
\label{sec:tool-demand}

LLM architecture determines which resources each pool consumes. Attention may be dense, sliding-window, or sparse, while its KV representation may use grouped-query attention (GQA), multi-query attention (MQA), or multi-head latent attention (MLA). FFN layers may be dense or MoE, with routed and shared experts. These choices change projection arithmetic, weight traffic, KV residency and read span, and the model state that must fit in each pool~\cite{pope2023efficiently,chen2022transformerarithmetic}. Input sequence length (ISL), resident context, output sequence length (OSL), batch, and precision scale those demands. \tool{} encodes the model through its layer, projection, attention, expert, context, and precision parameters. \tool{} counts causal attention until a sliding window caps it, uses the declared GQA/MQA or MLA dimensions, activates top-$k$ rather than all $E$ routed experts, and accounts for shared experts separately. Each hardware entry supplies the corresponding compute, bandwidth, capacity, network, and nameplate-power parameters. \tool{} derives compute and memory time from operator-level arithmetic and modeled memory traffic, then combines them with collective costs and frozen efficiency factors~\cite{williams2009roofline,hill2019gables}. \tool{} models prefill as one service phase over the full prompt. Its residency check includes weights, prompt KV, and one chunk of activations in usable memory. For MoE decode, \tool{} estimates attention and FFN service times separately and composes them with dispatch and combine costs. This demand--supply split lets \tool{} evaluate specification ranges for hardware that does not yet exist~\cite{shao2014aladdin}. For each pool, \tool{} models FLOPs, bytes, state, feasibility, service time, and token throughput.
\par
\endgroup
\endgroup

LLM architecture changes latency and feasibility through different paths. Dense attention makes KV residency and per-step reads grow with context, while sliding-window attention caps both. Sparse attention reduces selected-token reads but retains the full resident cache and index scan. Sparse attention can therefore reduce time without removing a capacity wall. GQA shards KV only through its supported head factor, whereas MLA KV remains replicated across tensor-parallel ranks. FFN service reads the dense-layer, routed-expert, and shared-expert weights used on each step. Uneven token routing can leave some expert-hosting units idle while the busiest unit finishes its work. This can lengthen FFN service and concentrate dispatch and combine traffic. \tool{} uses mean routed load in the main comparison and evaluates expert imbalance as a sensitivity (\S\ref{sec:discussion}). \tool{} charges every expert assigned to a unit against that unit's capacity. A pool is infeasible when its weights, KV state, and activations do not fit in usable memory. For feasible decode pools, \tool{} combines compute and memory time with partial overlap, then adds collectives and fixed per-layer costs. \tool{} allows decode microbatches to overlap, with pipeline fill and drain evaluated as a sensitivity (\S\ref{sec:robust}). \tool{} streams every modeled weight and KV byte from each hardware entry's serving memory tier.

\subsection{End-to-end serving metrics}
\label{sec:tool-metrics}

TTFT includes prefill service (\S\ref{sec:tool-demand}), the KV handoff, and waits for prefill and decode admission. The handoff transfers completed prompt KV state once per request and contributes to TTFT, not steady-state TPOT. We model it as store-and-forward without compression, prefix reuse, or layer-wise streaming.

\begin{figure}[!b]
\vspace{-16pt}
\centering
\begin{tikzpicture}[x=0.92cm, y=0.88cm, font=\scriptsize,
  lane/.style={fill=black!5, draw=none},
  mbone/.style={fill=tlslate, draw=black, line width=0.35pt},
  mbtwo/.style={fill=tlsand, draw=black, line width=0.35pt},
  adj/.style={dash pattern=on 1.4pt off 0.9pt},
  idle/.style={fill=white, postaction={pattern=north east lines, pattern color=tlbrick},
    draw=black, line width=0.35pt},
  lanelabel/.style={anchor=east, inner sep=0pt, text height=4.78pt, text depth=1.52pt},
  legendlabel/.style={anchor=west, inner sep=1pt, text height=4.78pt, text depth=1.52pt},
  paneltitle/.style={anchor=west, inner sep=0pt, font=\scriptsize\bfseries\boldmath},
  trip/.style={-{Latex[length=3.2pt,width=3.2pt]}, line width=0.7pt, black},
  dimL/.style={{Bar[width=3pt]Latex[length=2.2pt,width=2pt]}-, line width=0.45pt, draw=tlgray},
  dimR/.style={-{Latex[length=2.2pt,width=2pt]Bar[width=3pt]}, line width=0.45pt, draw=tlgray}
]
\definecolor{tlslate}{HTML}{3D5A80}%
\definecolor{tlsand}{HTML}{E6D2A4}%
\definecolor{tlbrick}{HTML}{A3402F}%
\definecolor{tlgray}{HTML}{4A4844}%
\def\bh{0.125}
\def\tlA#1#2#3#4{\path[mbone] (#2,#1-\bh) rectangle (#3,#1+\bh);
  \node[text=white, inner sep=0pt] at ({(#2+#3)/2},#1) {#4};}
\def\tlB#1#2#3#4{\path[mbtwo] (#2,#1-\bh) rectangle (#3,#1+\bh);
  \node[inner sep=0pt] at ({(#2+#3)/2},#1) {#4};}
\def\tlAd#1#2#3#4{\path[mbone,adj] (#2,#1-\bh) rectangle (#3,#1+\bh);
  \node[text=white, inner sep=0pt] at ({(#2+#3)/2},#1) {#4};}
\def\tlBd#1#2#3#4{\path[mbtwo,adj] (#2,#1-\bh) rectangle (#3,#1+\bh);
  \node[inner sep=0pt] at ({(#2+#3)/2},#1) {#4};}
\def\tlI#1#2#3{\path[idle] (#2,#1-\bh) rectangle (#3,#1+\bh);}
\def\tllane#1#2#3{\path[lane] (0,#1-\bh) rectangle (#2,#1+\bh);
  \node[lanelabel] at (-0.12,#1) {#3};}
\def\tldim#1#2#3#4{\node[inner sep=1pt] (tldimlabel) at ({(#1+#2)/2},#3) {#4};
  \draw[dimL] (#1,#3) -- (tldimlabel.west); \draw[dimR] (tldimlabel.east) -- (#2,#3);}

\begin{scope}[shift={(-1.55,0)}]
  \path[mbone] (0,-0.09) rectangle (0.30,0.09);
  \node[legendlabel] at (0.36,0) {microbatch 1};
  \path[mbtwo] (2.35,-0.09) rectangle (2.65,0.09);
  \node[legendlabel] at (2.71,0) {microbatch 2};
  \path[mbtwo,adj] (4.71,-0.09) rectangle (5.01,0.09);
  \node[legendlabel] at (5.07,0) {adjacent layer};
  \path[idle] (7.15,-0.09) rectangle (7.45,0.09);
  \node[legendlabel] at (7.51,0) {exposed wait};
\end{scope}

\node[paneltitle] at (-1.55,-0.40) {(a) P/D, co-located decode with two-batch overlap};
\def\ya{-0.74}\def\yb{-1.06}
\tllane{\ya}{6.8}{Decode pool}
\tllane{\yb}{6.8}{In-pool EP}
\tlA{\ya}{0}{0.75}{A1}   \tlB{\ya}{0.75}{1.5}{A2}
\tlA{\ya}{1.5}{2.45}{F1} \tlB{\ya}{2.45}{3.4}{F2}
\tlA{\ya}{3.4}{4.15}{A1} \tlB{\ya}{4.15}{4.9}{A2}
\tlA{\ya}{4.9}{5.85}{F1} \tlB{\ya}{5.85}{6.8}{F2}
\tlBd{\yb}{0}{0.45}{C2}
\tlA{\yb}{0.75}{1.2}{D1}  \tlB{\yb}{1.5}{1.95}{D2}
\tlA{\yb}{2.45}{2.9}{C1}  \tlB{\yb}{3.4}{3.85}{C2}
\tlA{\yb}{4.15}{4.6}{D1}  \tlB{\yb}{4.9}{5.35}{D2}
\tlA{\yb}{5.85}{6.3}{C1}
\tldim{0}{3.4}{-1.34}{layer $\ell$: $T/L$}
\tldim{3.4}{6.8}{-1.34}{layer $\ell{+}1$}

\node[paneltitle] at (-1.55,-1.72) {(b) P/D+AFD, ping-pong with $M=2$: exchange hidden};
\def\yA{-2.06}\def\yD{-2.38}\def\yF{-2.70}\def\yC{-3.02}
\tllane{\yA}{6.4}{Attn pool}
\tllane{\yD}{6.4}{A$\to$F link}
\tllane{\yF}{6.4}{FFN pool}
\tllane{\yC}{6.4}{F$\to$A link}
\tlA{\yA}{0}{1.6}{A1}   \tlB{\yA}{1.6}{3.2}{A2}
\tlA{\yA}{3.2}{4.8}{A1} \tlB{\yA}{4.8}{6.4}{A2}
\tlBd{\yD}{0}{0.35}{2}
\tlA{\yD}{1.6}{1.95}{1}  \tlB{\yD}{3.2}{3.55}{2}  \tlA{\yD}{4.8}{5.15}{1}
\tlBd{\yF}{0.35}{1.15}{F2}
\tlA{\yF}{1.95}{2.75}{F1} \tlB{\yF}{3.55}{4.35}{F2} \tlA{\yF}{5.15}{5.95}{F1}
\tlBd{\yC}{1.15}{1.5}{2}
\tlA{\yC}{2.75}{3.1}{1}  \tlB{\yC}{4.35}{4.7}{2}  \tlA{\yC}{5.95}{6.3}{1}
\draw[trip] (3.10,\yC+\bh) -- (3.10,\yA-\bh);
\tldim{0}{3.2}{-3.30}{layer $\ell$: $T/L$}
\tldim{3.2}{6.4}{-3.30}{layer $\ell{+}1$}

\node[paneltitle] at (-1.55,-3.68) {(c) Same, with a slower inter-pool link: exchange exposed};
\def\yA{-4.02}\def\yD{-4.34}\def\yF{-4.66}\def\yC{-4.98}
\tllane{\yA}{7.6}{Attn pool}
\tllane{\yD}{7.6}{A$\to$F link}
\tllane{\yF}{7.6}{FFN pool}
\tllane{\yC}{7.6}{F$\to$A link}
\tlA{\yA}{0}{1.6}{A1}   \tlB{\yA}{1.6}{3.2}{A2}   \tlI{\yA}{3.2}{3.8}
\tlA{\yA}{3.8}{5.4}{A1} \tlB{\yA}{5.4}{7.0}{A2}   \tlI{\yA}{7.0}{7.6}
\tlA{\yD}{1.6}{2.3}{1}  \tlB{\yD}{3.2}{3.9}{2}
\tlA{\yD}{5.4}{6.1}{1}  \tlB{\yD}{7.0}{7.6}{2}
\tlBd{\yF}{0.1}{0.9}{F2}
\tlA{\yF}{2.3}{3.1}{F1} \tlB{\yF}{3.9}{4.7}{F2} \tlA{\yF}{6.1}{6.9}{F1}
\tlBd{\yC}{0.9}{1.6}{2}
\tlA{\yC}{3.1}{3.8}{1}  \tlB{\yC}{4.7}{5.4}{2}  \tlA{\yC}{6.9}{7.6}{1}
\draw[trip] (3.80,\yC+\bh) -- (3.80,\yA-\bh);
\tldim{0}{3.8}{-5.26}{layer $\ell$: $T/L$}
\tldim{3.8}{7.6}{-5.26}{layer $\ell{+}1$}

\node[paneltitle] at (-1.55,-5.64) {(d) Fast link, busier FFN device: the FFN pool sets the period};
\def\yA{-5.98}\def\yF{-6.30}
\tllane{\yA}{7.2}{Attn pool}
\tllane{\yF}{7.2}{FFN pool}
\tlA{\yA}{0}{1.2}{A1}   \tlI{\yA}{1.2}{1.8} \tlB{\yA}{1.8}{3.0}{A2} \tlI{\yA}{3.0}{3.6}
\tlA{\yA}{3.6}{4.8}{A1} \tlI{\yA}{4.8}{5.4} \tlB{\yA}{5.4}{6.6}{A2} \tlI{\yA}{6.6}{7.2}
\tlBd{\yF}{0}{1.55}{F2}
\tlA{\yF}{1.55}{3.35}{F1} \tlB{\yF}{3.35}{5.15}{F2} \tlA{\yF}{5.15}{6.95}{F1}
\tlB{\yF}{6.95}{7.2}{}
\tldim{0}{3.6}{-6.58}{layer $\ell$: $T/L$}
\tldim{3.6}{7.2}{-6.58}{layer $\ell{+}1$}
\node[inner xsep=2pt, inner ysep=0pt, text=tlgray] (tltime) at (3.8,-6.84) {time};
\draw[line width=0.45pt, draw=tlgray] (0,-6.84) -- (tltime.west);
\draw[-{Latex[length=2.2pt,width=2.2pt]}, line width=0.45pt, draw=tlgray]
  (tltime.east) -- (7.6,-6.84);
\end{tikzpicture}
\vspace{-4pt}
\caption{\textbf{Microbatch overlap hides the attention/FFN exchange only when the busier pool sets the steady-state period.} Steady-state decode schedules over two layers with $M=2$ microbatches. A, D, F, C: attention, dispatch, FFN, combine. Arrows mark microbatch~1's return. (a)~Co-located decode hides D and C. (b)~AFD ping-pong hides the exchange in the slack of the shorter FFN stage. Balanced stages need $M\geq 3$ (\S\ref{sec:rolehw}). (c)~A slower link's round trip outlasts both microbatches' attention; the attention pool idles. (d)~A busier FFN device lengthens F to set the period; attention waits. Routed tokens can crowd one device's experts.}
\label{fig:bg-timeline}
\end{figure}
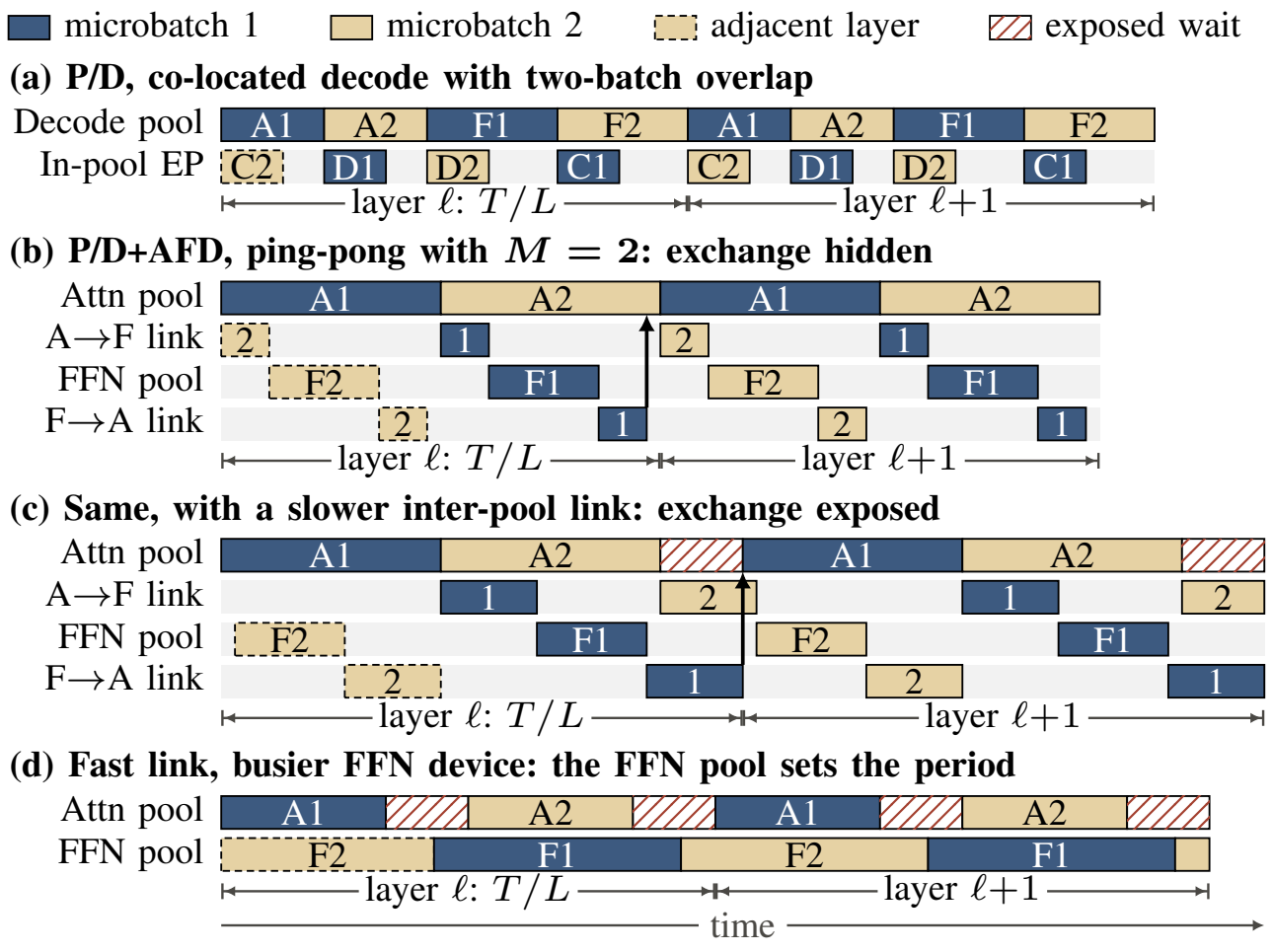

TPOT depends on the recurring decode schedule. Figure~\ref{fig:bg-timeline} illustrates how overlap affects TPOT through several example decode schedules. The attention/FFN boundary exchanges activations twice per MoE layer of every decode step. It lengthens TPOT unless microbatch overlap hides the exchange. Co-located decode can hide its in-pool dispatch and combine with \emph{two-batch overlap} (\figpanel{fig:bg-timeline}{a}). One microbatch then communicates while the other computes~\cite{deepseek2025inferencesys,sglang2025largeep}. P/D+AFD hides the exchange with \emph{ping-pong} pipelining of $M$ microbatches between the pools~\cite{zhu2025megascaleinfer,fu2026fastafd}. With $M$ microbatches of $b$ requests in flight, each resource serves the phases in one fixed order. TPOT is this schedule's steady \emph{period}~$T$, in which every resident request emits one token. For a homogeneous P/D+AFD stack ($M\le 8$) with a dedicated inter-pool link,
\par
\begingroup
\fontsize{9}{11}\selectfont
\setlength{\abovedisplayskip}{-3pt}
\setlength{\abovedisplayshortskip}{-3pt}
\setlength{\belowdisplayskip}{11pt}
\setlength{\belowdisplayshortskip}{11pt}
\begin{equation}\label{eq:period}
T = L\max\!\big(M\max(t_{\mathrm{A}},t_{\mathrm{F}},t_{\mathrm{D}},t_{\mathrm{C}}),\ t_{\mathrm{A}}+t_{\mathrm{D}}+t_{\mathrm{F}}+t_{\mathrm{C}}\big),
\end{equation}
\endgroup
where $t_{\mathrm{A}}$, $t_{\mathrm{D}}$, $t_{\mathrm{F}}$, and $t_{\mathrm{C}}$ are one microbatch's attention, dispatch, FFN, and combine times in each of $L$ layers. The busier pool sets $T$ in \figpanel{fig:bg-timeline}{b} and \figpanel{fig:bg-timeline}{d}, hiding the exchange (\S\ref{sec:rolehw}). In \figpanel{fig:bg-timeline}{c}, the round trip $t_{\mathrm{A}}+t_{\mathrm{D}}+t_{\mathrm{F}}+t_{\mathrm{C}}$ sets $T$ and exposes the exchange. A link can also set $T$. Max-plus algebra exactly evaluates this order for mixed dense and MoE stacks and co-located P/D ($M\le 2$). A shared link also carries the in-pool collectives. A design holds $U=Mb$ resident requests and delivers $X=U/T$ tokens/s. With $M=1$, Eq.~\eqref{eq:period} is the submitted serial law. The fill-and-drain effect remains a sensitivity (\S\ref{sec:robust}).

\begingroup
\clubpenalty=10000
SLO goodput is the largest request arrival rate satisfying both TTFT and TPOT targets and the stability constraint (\S\ref{sec:model-equations}). At each arrival rate, \tool{} adds the modeled 99th-percentile (p99) of summed prefill and decode queue waits to base TTFT and tests these constraints. Pool separation can remove interference and enable role-specific batching, but it adds queues and caps SLO goodput at that of the least-admissible pool. \tool{} returns SLO goodput and the \emph{saturation throughput} of the complete deployment, its largest sustainable token rate. A design can gain throughput yet lose goodput by lengthening TPOT near a tight target. We therefore report both and vary the target (\S\ref{sec:context}). By default, both designs use the same nameplate power ceiling. We report goodput per kW using the selected hardware's summed rated power, reflecting how many accelerators a power-capped rack may admit~\cite{fan2007provisioning,patel2024polca}. Decode-only comparisons use decode kW. We also normalize goodput by \emph{deployment unit}, the smallest unit a vendor ships (\S\ref{sec:inputs-hw}). We include only mechanisms with explicit cost models and call one deployment better only when its margin passes the two-part test (\S\ref{sec:cal-test}).
\par
\endgroup

%% file: sections/3-Inputs.tex
\section{Evaluation Scenarios}\label{sec:inputs}\label{sec:design}

\subsection{Serving decomposition and hardware assignment}
\label{sec:inputs-axes}

Our main comparison evaluates the incremental attention/FFN split by comparing \textbf{P/D} with \textbf{P/D+AFD} (Figure~\ref{fig:inputs-design-axes}). Unified serving is considered separately in the KV-footprint sensitivity (\S\ref{sec:context}), where its modeled SLO goodput is an upper bound because chunked prefill is hidden in decode steps. A \emph{placement policy} constrains hardware assignment by specifying which pools may use different accelerator classes. \textbf{One-class placement} uses one class for every role and compares homogeneous deployments. \textbf{Phase-split} lets prefill and decode differ. \textbf{Role-specific} lets prefill, attention, and FFN differ. \textbf{FFN-targeted} holds prefill and attention on one class and chooses the FFN class freely.

\input{table/hardware_catalog_status}
\input{table/class_ledger}

\subsection{Four accelerator classes expose different resource tradeoffs}
\label{sec:inputs-hw}

\tool{}'s hardware catalog contains 39 accelerators across six archetypes, with current, announced, and speculative entries (Table~\ref{tab:hardware}). Archetypes group hardware by organization. Table~\ref{tab:hardware} lists the accelerators, their release status, and their study roles. Commodity DRAM and model-specific ASIC appear in the catalog sweep but not the main comparison.

\begin{figure*}[t]
  \centering
  \includegraphics[width=\textwidth,height=2.30in,keepaspectratio]{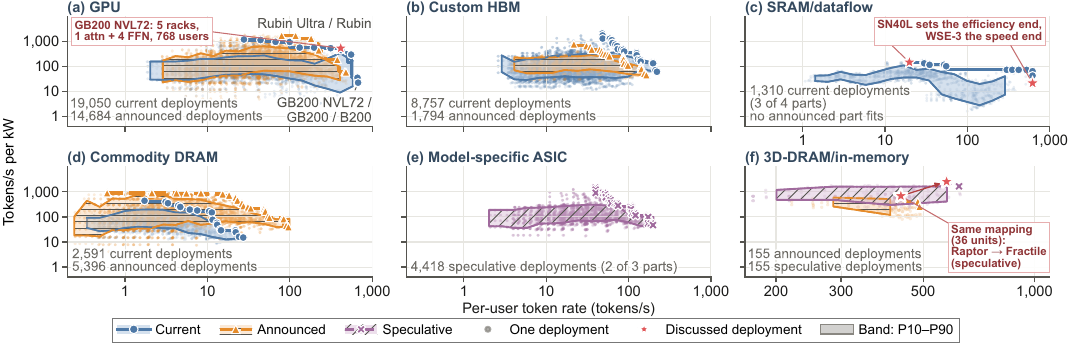}
  \vspace{-22pt}
  \caption{\textbf{Coverage of the hardware catalog under one controlled workload.} (a)--(f)~Six
  archetypes; DeepSeek-V3 deployments with ISL and OSL both 16K, one-class placement for decode attention
  and FFN, $\leq$128 units, and $\leq$1\,MW; 36 of 39 accelerators support at least one of the 58,310
  feasible deployments. Parts without native FP8 run FP8-stored weights at their 16-bit rate. Bands:
  P10--P90 of deployments; \figpanel{fig:inputs-catalog-coverage}{f} has its own x range.}
  \label{fig:inputs-catalog-coverage}
  \vspace{-15pt}
\end{figure*}
\input{table/model_catalog_status}

Using accelerator anchors from Table~\ref{tab:hardware}, we define four prospective classes for the main comparison: GPU, systolic, 3D-DRAM, and SRAM/dataflow. These classes span different resource tradeoffs because accelerator roadmaps do not scale compute, memory bandwidth, and capacity together~\cite{nvidia2026rubinspecs,dmatrix2025alchip3ddram,nvidia2026groq3lpx,google2026tpu7x}. TPU~v7's memory system anchors the constructed systolic class, although Table~\ref{tab:hardware} groups that accelerator under Custom HBM. \tool{} models each class's resources as \emph{ranges} to test comparisons across uncertainty in published and estimated specifications. Table~\ref{tab:ledger} lists the nominal resource values and declared ranges for each class. The 32 \emph{specification corners} (64 with the photonic band) combine endpoints of compute, memory bandwidth, capacity, links, and power. At each corner, we optimize each design's allowed hardware assignments, mappings, and pool sizes for maximum SLO goodput within the common power budget. For each resource, all classes use their own lower or upper endpoint together. Lower-confidence inputs use wider ranges, and no capability corner falls below its anchor. A range-width curve scales every widening factor from $0.5\times$ to $2\times$.

\begingroup
\setlength{\parskip}{4pt}
\looseness=1
Figure~\ref{fig:inputs-catalog-coverage} illustrates \tool{}'s mapping coverage across the full hardware catalog for DeepSeek-V3 with ISL and OSL both 16K. The sweep uses one-class placement for decode attention and FFN, with ceilings of 128 units and 1\,MW. It varies the hardware system, release status, mapping (TP, DP, EP, batch, and $M\leq8$), and deployment size. Moving right means faster token delivery to each user; moving up means more throughput per kW. Each frontier traces feasible mapping and deployment choices. The GB200 NVL72 star in \figpanel{fig:inputs-catalog-coverage}{a} marks a balanced deployment with $4\times$ as many FFN racks as attention racks. Points farther right trade throughput per kW for faster token delivery by serving fewer users. \figpanel{fig:inputs-catalog-coverage}{b} covers custom-HBM accelerators. In \figpanel{fig:inputs-catalog-coverage}{c}, SN40L sets the efficiency end of the current SRAM/dataflow frontier, and WSE-3 sets the speed end. The preferred system depends on the deployment target. \figpanel{fig:inputs-catalog-coverage}{d} and \figpanel{fig:inputs-catalog-coverage}{e} cover commodity-DRAM and model-specific ASIC accelerators, respectively. Groq~3 LPX and Groq LPU v1 in \figpanel{fig:inputs-catalog-coverage}{c} cannot hold the model in 128 units, and Taalas HC1 in \figpanel{fig:inputs-catalog-coverage}{e} targets another model. The matched mapping in \figpanel{fig:inputs-catalog-coverage}{f} predicts $3.7\times$ Raptor's throughput per kW for Fractile, using projected Fractile and \mbox{announced Raptor specifications.}
\par
\endgroup

\begingroup
\setlength{\parskip}{0pt}
\renewcommand{\subsection}{\csname @startsection\endcsname{subsection}{2}{0pt}{1.5ex plus 1.5ex minus 0.5ex}{0.7ex}{\normalfont\normalsize\itshape}}
\subsection{Communication fabrics and collectives}
\label{sec:inputs-fabric}

Optical fabrics are maturing across the interconnect hierarchy~\cite{nvidia2025photonics,ayarlabs2025teraphy,lightmatter2026passage3d}. We model electrical links as the current baseline and use the photonic roadmap range listed in Table~\ref{tab:ledger} for sensitivity analysis. \tool{} represents each fabric as a hierarchy of link tiers defined by bandwidth, latency, and supported group size. Within each fabric, we compare a dedicated inter-pool link with a shared scale-up domain, where the attention/FFN exchange competes with in-pool collectives (e.g., local TP reduction~\cite{nvidia2026nvlinkpage}). A faster inter-pool link shortens AFD's exchange, but the same upgrade in a shared domain also accelerates P/D's in-pool collectives. We evaluate fabric and serving choices jointly under both modes.\par
\setlength{\parskip}{0pt}
\looseness=1
Data movement and synchronization can erase the throughput benefit of additional TP, PP, DP, or EP~\cite{pope2023efficiently,narayanan2021megatron,lepikhin2021gshard,shen2026collectives}. \tool{} derives message sizes from the model, mapping, and precision, and sizes DP replicas. It charges the expert all-to-all per hop on the tier that hop crosses (\S\ref{sec:model-equations}) and every other collective on the tier that spans its participants. \tool{} models in-network collective reduction only when the selected fabric supports it (e.g., SHARP~\cite{nvidia2026sharp}). Otherwise, it uses recursive halving/doubling (Rabenseifner), with ring and offload-disabled \mbox{implementations as sensitivities.}
\par
\endgroup

\subsection{LLM architecture directions}
\label{sec:inputs-models}

\tool{} evaluates specific LLM models and controlled future directions. These vary model properties separately rather than along one ``model size,'' following recent frontier families that move expert width, attention state, and context on separate axes~\cite{deepseek2026v4,minimax2026m3,qwen2026qwen35,moonshot2026k26}. Expert-heavy MoE increases FFN work and expert communication while activating only a fraction of its weights~\cite{shazeer2017moe,lepikhin2021gshard,fedus2022switch}. Total weight capacity and active compute can therefore grow differently. Attention variants, from GQA~\cite{ainslie2023gqa} to MLA~\cite{deepseek2024v2}, change the KV bytes stored per token. Longer contexts make this state a capacity and data-movement constraint~\cite{kwon2023pagedattention,qin2025mooncake,yu2026amma}. Sparse attention reduces selected KV reads per step while retaining the complete cache~\cite{deepseek2025nsa,lu2025moba,deepseek2025v32}. A model can read little KV yet exceed attention-pool capacity.

The model catalog contains 26 entries, including 25 published model anchors and one scaled 2T MoE proxy. Table~\ref{tab:models}(b) lists 20 of these anchors and the proxy. The main comparison uses four controlled future directions rather than averaging across the catalog. Each compound synthetic scenario extends one model family mainly along one demand dimension~\cite{clark2022unified} while rescaling width or depth, as shown in Table~\ref{tab:models}(a). Their resident contexts are $C=8$K, 32K, and 128K with $\mathrm{ISL}=C$.
Model mechanisms enter through KV bytes per token and bytes read per step. Gated attention adds no KV state~\cite{qiu2025gatedattention}. A hybrid with full attention in a fraction $f$ of layers, such as Qwen3.5 ($f=1/4$)~\cite{qwen2026qwen35}, stores about $f$ times the KV per token plus a fixed state per request. That storage sets its capacity need.

\subsection{SLOs, goodput, workload, and budget}
\label{sec:inputs-objectives}

\tool{} sweeps the request arrival rate to test both latency targets (\S\ref{sec:model-equations}). In the main sweep, TPOT is fixed at each deployment's full-occupancy value, while queueing raises tail TTFT until its target fails. The separate load simulation lets TPOT vary with batch occupancy (Figure~\ref{fig:load-curves}). Each comparison \emph{cell} fixes one model direction, fabric, link-sharing mode, context, and placement policy. Its SLO pairs a TTFT target of 1 or 5\,s with a TPOT target of 10, 20, 50, or 200\,ms. The separate regime map sweeps eight contexts (4K--256K) and 24 TPOT targets (10--300\,ms) at nominal specifications. In the main workload, the prompt fills the resident context ($\mathrm{ISL}=C$), and the KV handoff moves $C$ tokens. Each request produces 1,024 output tokens and has no prefix-cache hit. Every policy uses a 128\,kW budget. Scale runs use 512\,kW. One-class placement is also evaluated at 64 units.
Under the main search rule, EP starts at 8 and doubles for both designs. We sweep $M=1,\ldots,8$ for P/D+AFD and $M=1,2$ for P/D. Prefill TP exceeds 8 only if no smaller replica fits.

The 16 \emph{headline comparisons} cross four directions, two fabrics, and two link-sharing modes. Each uses $C=32$K, a 50\,ms TPOT target, a 5\,s TTFT target, a 128\,kW budget, and FP8 weights, KV, and activations, except 16-bit combine. We define two verdict sets before presenting the results (\S\ref{sec:cal-test}). The \emph{best-allowed} set represents the architect's decision and gives each design its best allowed placement policy. The \emph{one-class} set gives each design its best single class. Both designs select the same class in 14 of the 16 comparisons.

%% file: table/hardware_catalog_status.tex
\begin{table*}[t]
\centering
\caption{Hardware catalog and study roles: 39 accelerators in six archetypes.}
\vspace{-7pt}
\label{tab:hardware}\label{tab:inputs-hw}
\scriptsize
\setlength{\tabcolsep}{3pt}
\renewcommand{\arraystretch}{1.0}
\newcommand{\RL}{\raggedright\let\\\tabularnewline}
\definecolor{figann}{HTML}{E58B2A}\colorlet{tierann}{figann!70!black}%
\definecolor{figspec}{HTML}{A06A9D}\colorlet{tierspec}{figspec!75!black}%
\newcommand{\tierA}[1]{#1}%
\newcommand{\tierS}[1]{#1}%
\begin{tabular*}{\textwidth}{@{\extracolsep{\fill}}p{0.15\textwidth}!{\tablecolsep}p{0.60\textwidth}!{\tablecolsep}p{0.21\textwidth}@{}}
\toprule
\RL Hardware archetype (entries) & \RL Catalog entries by release status\newline (unmarked: current; \tierA{\emph{Ann.}: announced}, pre-release; \tierS{\emph{Spec.}: speculative}, projected specifications) & \RL \raisebox{-0.5\baselineskip}[0pt][0pt]{Role in this paper} \\
\midrule
\RL GPU (13) & \RL A100~\cite{nvidia2020a100}; H100/H200~\cite{choquette2023hopper,nvidia2026h200};
  B200/GB200/GB200 NVL72~\cite{tirumala2024blackwell,nvidia2026gb200nvl72};
  AMD MI300X/MI325X/MI355X~\cite{amd2026acceleratorspecs}.
  \tierA{\emph{Ann.:} B100}~\cite{tirumala2024blackwell}; \tierA{GB300}~\cite{nvidia2025gb300};
  \tierA{Rubin}~\cite{nvidia2026rubinspecs}; \tierA{Rubin Ultra}~\cite{nvidia2026rubinultra}. &
  \RL GPU class (Rubin anchor) \\
\RL Custom HBM (11) & \RL Inferentia2~\cite{aws2023inf2}; Trainium2~\cite{aws2024trainium2};
  TPU v5e/v5p/v6e~\cite{google2023tpuv5e,google2023tpuv5p,google2024tpuv6e};
  TPU v7 Ironwood~\cite{google2026tpu7x};
  Gaudi~3~\cite{intel2024gaudi3}; Maia~100~\cite{microsoft2024maia100};
  RNGD~\cite{furiosa2024rngd}; ATOM~\cite{rebellions2023atom}.
  \tierA{\emph{Ann.:} REBEL}~\cite{rebellions2024rebel}. &
  \RL Systolic class (TPU~v7 anchor) \\
\RL 3D-DRAM/in-memory (2) & \RL \tierA{\emph{Ann.:} d-Matrix Raptor}~\cite{nair2026raptor}.
  \tierS{\emph{Spec.:} Fractile Transformer Chip}~\cite{fractile2026}. &
  \RL 3D-DRAM class (Raptor anchor) \\
\RL SRAM/dataflow (5) & \RL Cerebras WSE-3~\cite{cerebras2024wse3};
  Groq LPU v1~\cite{abts2022groq}; SambaNova SN40L~\cite{prabhakar2024sn40l};
  Untether speedAI240~\cite{untether2024speedai240}.
  \tierA{\emph{Ann.:} Groq~3 LPX}~\cite{nvidia2026groq3lpx}. &
  \RL SRAM/dataflow class (Groq~3 LPX anchor) \\
\midrule
\RL Commodity DRAM (5) & \RL MTIA v1/v2~\cite{meta2023mtia,meta2024mtia}; Tenstorrent
  Blackhole/Wormhole~\cite{tenstorrent2026blackhole,tenstorrent2026wormhole}.
  \tierA{\emph{Ann.:} Positron Asimov}~\cite{positron2026asimov}. & \RL Tool support (Figure~\ref{fig:inputs-catalog-coverage}) \\
\RL Model-specific ASIC (3) & \RL \tierS{\emph{Spec.:}
  Etched Sohu}~\cite{etched2024sohu}; \tierS{MatX C1}~\cite{matx2026one};
  \tierS{Taalas HC1}~\cite{taalas2026hc1}. & \RL Tool support (Figure~\ref{fig:inputs-catalog-coverage}) \\
\bottomrule
\end{tabular*}
\vspace{-11pt}
\end{table*}

%% file: table/class_ledger.tex
\begin{table*}[t]
\centering
\caption{Prospective accelerator classes modeled in this study. Values are nominal per deployment unit [lowest--highest specification corner].
Across classes, nominal values span less than $2\times$ in compute, $163\times$ in memory bandwidth, and $72\times$ in capacity.}
\label{tab:ledger}\label{tab:classes}
\vspace{-7pt}
\scriptsize
\setlength{\tabcolsep}{1.3pt}
\renewcommand{\arraystretch}{1.0}
\newcommand{\RL}{\raggedright\let\\\tabularnewline}
\newcommand{\rg}[1]{{\color{black!55}#1}}
\begin{tabular*}{\textwidth}{@{\extracolsep{\fill}}l!{\tablecolsep}l!{\tablecolsep}l!{\tablecolsep}%
  r@{\extracolsep{0pt}\,}l!{\extracolsep{\fill}\tablecolsep}r@{\extracolsep{0pt}\,}l!{\extracolsep{\fill}\tablecolsep}%
  r@{\extracolsep{0pt}\,}l!{\extracolsep{\fill}\tablecolsep}r@{\extracolsep{0pt}\,}l!{\extracolsep{\fill}\tablecolsep}%
  r!{\tablecolsep}r@{\extracolsep{0pt}\,}l!{\extracolsep{\fill}\tablecolsep}r@{\extracolsep{0pt}\,}l@{}}
\toprule
Class (anchor, memory) & Unit & Tier$^{1}$ & \multicolumn{2}{c!{\tablecolsep}}{FP8 PFLOPS\,$^{2}$} &
  \multicolumn{2}{c!{\tablecolsep}}{Mem.\ TB/s} & \multicolumn{2}{c!{\tablecolsep}}{Cap.\ GiB} &
  \multicolumn{2}{c!{\tablecolsep}}{Scale-up TB/s} & Domain & \multicolumn{2}{c!{\tablecolsep}}{Scale-out Tb/s} &
  \multicolumn{2}{c@{}}{Power W (est.)$^{3}$} \\
\midrule
\textbf{GPU}: Rubin$^{4}$~\cite{nvidia2026rubinspecs,nvidia2026nvlinkpage}, HBM4 & 2-die package & Ann.\ 1.25 &
  17.5 & \rg{[17.5--37.2]} & 19.2 & \rg{[19.2--30.0]} & 288 & \rg{[288--450]} &
  1.5 & \rg{[1.5--2.3]} & 72 & 1.6 & \rg{[1.6--2.5]} &
  1{,}800 & \rg{[1{,}665--2{,}475]} \\
\textbf{Systolic}$^{5}$: TPU~v7~\cite{google2026tpu7x}, HBM3e & 2-die chip & Datash.\ 1.0 &
  13.8 & \rg{[9.2--18.5]} & 7.4 & \rg{[7.4--11.1]} & 192 & \rg{[192--288]} &
  0.6 & \rg{[0.6--0.9]} & 9{,}216 & 0.10 & \rg{[0.10--0.15]} & 600 & \rg{[555--825]} \\
\textbf{3D-DRAM}: Raptor~\cite{nair2026raptor} & 2-chip card & Est.\ 1.5 &
  10.0 & \rg{[10.0--16.3]} & 100 & \rg{[100--225]} & 32 & \rg{[32--152]} &
  1.0 & \rg{[1.0--1.6]} & 8 & 0.8 & \rg{[0.8--1.3]}$^{6}$ & 844 & \rg{[591--1{,}097]} \\
\textbf{SRAM/dataflow}: Groq~3 LPX~\cite{nvidia2026groq3lpx} & 8-LPU tray & Ann.\ 1.25 &
  9.6 & \rg{[9.6--20.4]} & 1{,}200 & \rg{[1{,}200--2{,}550]} & 4.0 & \rg{[4.0--17.5]} &
  10.0 & \rg{[10.0--21.3]} & 32 & 0.8 & \rg{[0.8--1.7]}$^{6}$ & 3{,}000 & \rg{[2{,}100--3{,}900]} \\
\midrule
\multicolumn{16}{@{}p{\textwidth}@{}}{\RL \textbf{Electrical fabric:} Class
  tiers as above, per direction and unit (vendor bidirectional figures halved). Per-hop
  latency is 0.5--2\,\textmu s on scale-up tiers and an estimated 4\,\textmu s on scale-out, with
  4--8\,\textmu s as a sensitivity~\cite{hamidouche2025gin}. A cross-class inter-pool link runs at the
  slower endpoint's scale-out rate (systolic data-center network, DCN: 0.1\,Tb/s = 12.5\,GB/s).} \\
\multicolumn{16}{@{}p{\textwidth}@{}}{\RL \textbf{Photonic fabric:} At nominal specifications, every
  inter-unit tier both designs use runs at no less than 12\,Tb/s per direction
  \rg{[4--32]}, with latency as electrical. Public optical I/O parts deliver
  8--64\,Tb/s (both directions) per engine~\cite{ayarlabs2025teraphy,lightmatter2025l200}
  and 114\,Tb/s per interposer~\cite{lightmatter2025m1000}.} \\
\multicolumn{16}{@{}p{\textwidth}@{}}{\RL $^{1}$Confidence and range-widening factor; Datash.: datasheet; Ann.: announced; Est.: estimated.
  $^{2}$Dense peaks where published. $^{3}$Nameplate, without hosts, switches, or cooling.
  $^{4}$Rubin HBM and NVLink values differ across NVIDIA
  sources~\cite{nvidia2026verarubinds,aubrey2026verarubin,nvidia2026rubinspecs}. The range covers both.
  $^{5}$Constructed from the TPU~v7 memory system with $2$--$4\times$ its 4.6\,PFLOPS
  ($3\times$ nominal; not widened). $^{6}$Estimate. 3D-DRAM also uses a PCIe~7 rack scale-up tier
  at 0.256\,TB/s per direction, with an estimated 64-card domain.} \\
\bottomrule
\end{tabular*}
\vspace{-12pt}
\end{table*}

%% file: table/model_catalog_status.tex
\begin{table*}[t]
\centering
\caption{Prospective directions and model catalog. The four directions span 35--55B active parameters
($1.6\times$). Total weights vary by $5.6\times$, and stored KV per token varies by $8.9\times$.
The directions use $C=8$K/32K/128K ($\mathrm{ISL}=C$, $\mathrm{OSL}=1{,}024$, FP8). KV per request is KV per token
$\times(C+\mathrm{OSL})$. A step reads KV per token $\times\,C$. Sparse attention instead reads the
top-2{,}048 tokens plus the full index.}
\label{tab:models}\label{tab:inputs-models}
\vspace{-6pt}%
\scriptsize
\setlength{\tabcolsep}{2pt}
\renewcommand{\arraystretch}{1.0}
\newcommand{\RL}{\raggedright\let\\\tabularnewline}
\newcommand{\RLh}{\RL\hangindent=0.45em\hangafter=1\relax}
\newcommand{\lightcmid}[3]{\noalign{\xdef\BTsvA{\the\aboverulesep}\xdef\BTsvB{\the\belowrulesep}%
    \global\aboverulesep=#1\relax\global\belowrulesep=#2\relax}%
  \cmidrule#3%
  \noalign{\global\aboverulesep=\BTsvA\relax\global\belowrulesep=\BTsvB\relax}}
\newcommand{\pb}[1]{\multicolumn{2}{@{\hspace{\tabcolsep}}p{\dimexpr0.42\textwidth+2\tabcolsep\relax}}{\RL #1}}
\newcommand{\pbe}[1]{\multicolumn{2}{@{\hspace{\tabcolsep}}p{\dimexpr0.42\textwidth+2\tabcolsep\relax}@{}}{\RL #1}}
\begin{tabular}{@{}p{\dimexpr\textwidth-0.84\textwidth-8\tabcolsep\relax}p{0.21\textwidth}p{0.21\textwidth}p{0.21\textwidth}p{0.21\textwidth}@{}}
\toprule
\textbf{(a)} & \multicolumn{4}{@{\hspace{\tabcolsep}}l@{}}{\textbf{Four prospective model directions (nominal)}} \\
\lightcmid{1pt}{1.2pt}{(l{\tabcolsep}){2-5}}
 & \RL\textbf{Expert-heavy MoE} & \RL\textbf{KV-optimized MLA} &
  \RL\textbf{KV-heavy GQA-MoE} & \RL\textbf{Sparse-attention MLA} \\
\RL \textbf{Demand varied} & \RL \textbf{FFN work, expert capacity} &
  \RL \textbf{KV bytes per token} & \RL \textbf{KV-cache capacity} &
  \RL \textbf{KV reads per step} \\
\midrule
\RL Anchor (total/active) & \RL Kimi~K2~\cite{kimi2025k2} (1.04T/32B) &
  \RL DeepSeek-V3~\cite{deepseek2024v3} (671B/37B) &
  \RL Qwen3-235B~\cite{qwen2025qwen3} (235B/22B) &
  \RL DeepSeek-V3.2-Exp~\cite{deepseek2025v32} (685B/37B) \\
\RL Trend from\newline (published models) &
  \RLh 3: DeepSeek-V2, V3, Kimi~K2~\cite{deepseek2024v2,deepseek2024v3,kimi2025k2}:
  experts 160$\rightarrow$256$\rightarrow$384, active share
  8.9$\rightarrow$5.5$\rightarrow$3.1\% &
  \RLh 4: DeepSeek-V2, V3, R1, Kimi~K2~\cite{deepseek2024v2,deepseek2024v3,deepseek2025r1,kimi2025k2}:
  latent rank 512 in all four; a what-if &
  \RLh 1: Qwen3-235B's 96\,kB/token FP8 GQA cache; the size range is a scenario &
  \RLh 1: V3.2-Exp sparse attention~\cite{deepseek2025v32,deepseek2025nsa}:
  top-2{,}048 tokens, 128-wide index \\
\RL Construction & \RLh Experts $\times3$, width $\times1.2$, expert FFN
  $\times1.125$; active share 1.0\% &
  \RLh Latent rank 512$\rightarrow$320, width $\times1.1$ &
\RLh Width $\times1.31$, depth $\times1.73$ (94$\rightarrow$163 layers), every layer MoE &
  \RLh Width $\times1.1$; budget and index fixed \\
\lightcmid{1pt}{1.2pt}{(l{\tabcolsep}){2-5}}
\RL Total / active params & 4.11T / 42B & 737B / 35B & 916B / 55B & 737B / 35B \\
\RL Layers / MoE layers & 61 / 60 & 61 / 58 & 163 / 163 & 61 / 58 \\
\RL Experts rt./sh./top-$k$ & 1{,}152 / 1 / 8 & 256 / 1 / 8 & 168 / 0 / 8 & 256 / 1 / 8 \\
\RL Width / query heads & 8{,}576 / 64 & 7{,}936 / 64 & 5{,}376 / 40 (5 KV) & 7{,}936 / 64 \\
\RL KV per token (kB) & 35.1 & 23.4 & 208.6 & 42.9 (35.1 + index 7.8) \\
\bottomrule
\end{tabular}

\vspace{1pt}
\begin{tabular}{@{}p{\dimexpr\textwidth-0.84\textwidth-8\tabcolsep\relax}p{0.21\textwidth}p{0.21\textwidth}p{0.21\textwidth}p{0.21\textwidth}@{}}
\RL \textbf{(b)} Family & \pb{\textbf{Catalog}: 20 published model anchors plus one scaled proxy} &
  \pbe{Role in the study} \\
\midrule
\RL Dense transformer (6) & \pb{Llama-3.1 8B/70B/405B~\cite{grattafiori2024llama3}; Mistral 7B; OPT-13B;
  Qwen2.5-72B} & \pbe{Calibration and controls; no synthetic dense growth point} \\
\RL MoE with GQA (9) & \pb{Mixtral 8x7B/8x22B~\cite{jiang2024mixtral}; Qwen3 30B/235B; Llama 4
  Scout/Maverick; GPT-OSS 20B/120B~\cite{openai2025gptoss}; Grok-1} & \pbe{Uncompressed GQA KV; Qwen3-235B
  seeds the KV-heavy direction} \\
\RL MoE with MLA (5+1) & \pb{DeepSeek-V2/V3/R1/V3.2-Exp; Kimi K2; scaled 2T MoE
  proxy} & \pbe{Latent KV; seeds the other three directions} \\
\bottomrule
\end{tabular}
\vspace{-15pt}%
\end{table*}

%% file: sections/4a-Insights.tex
\section{When Does Splitting Decode Pay?}
\label{sec:insights}

We vary the serving decomposition and hardware assignment independently,
in four steps. Does adding the attention/FFN boundary help under a common budget? Which parts of
decoded-token latency, and which link, determine whether role-specific
placement repays that boundary? When a fixed deployment exceeds capacity,
which lever restores feasibility? Which hardware assignments remain stable as
the model, fabric, hardware assumptions, and objective change?

\subsection{Under one budget, the ranking depends on pool allocation}
\label{sec:regime}\label{sec:insights-2x2}

\begin{figure*}[!t]
\vspace{-10pt}
\centering
\includegraphics[width=\textwidth,height=1.62in]{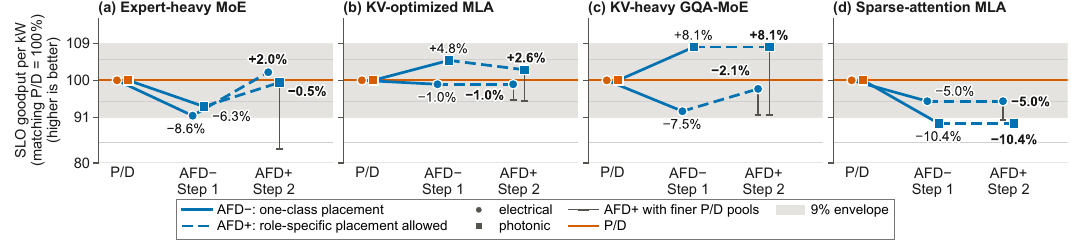}
\vspace{-23pt}
\caption{\textbf{Under the revised model and 128\,kW,
adding the attention/FFN boundary on one class moves SLO goodput per kW by
$-10.4\%$ to $+8.1\%$, and role-specific placement changes little. Seven of
eight \mbox{AFD+} margins lie inside the 9\% envelope.}
(a)--(d)~Headline cells (32K, TPOT 50\,ms, TTFT 5\,s, dedicated). Whiskers allow any integer EP for P/D ($-16.6\%$ to $+2.9\%$ under \mbox{AFD$+$}).}
\label{fig:insights-2x2}
\vspace{-12pt}
\end{figure*}

Does the attention/FFN split help by itself, or through different accelerator classes? Figure~\ref{fig:insights-2x2}
compares the four model directions relative to P/D's SLO goodput per kW at
100\%. \mbox{AFD$-$} (Step~1) compares each design's best one-class
deployment. \mbox{AFD$+$} (Step~2) allows role-specific placement for
P/D+AFD and phase-split placement for P/D. Circles mark electrical cases
and squares photonic. Solid and dashed blue segments lead to \mbox{AFD$-$}
and \mbox{AFD$+$}, respectively. The gray band marks the 9\% envelope around
P/D. Whiskers show \mbox{AFD$+$} with finer P/D decode pool sizes.

Each comparison independently searches both designs' mappings and pool sizes for
the highest feasible SLO goodput under 128\,kW, reported per kW. Workload, SLO,
precision, calibration, and available classes stay fixed. Each comparison uses
its own optimized P/D reference. At nominal values, the \mbox{AFD$-$}
comparison has no consistent winner. Role-specific placement in \mbox{AFD$+$} changes little.
Under the main power-of-two EP rule, only photonic sparse attention lies
beyond the 9\% envelope (favoring P/D), and none is robust over the tested
specification corners (\S\ref{sec:decide}).

\begin{figure*}[!b]
  \vspace{-10pt}
  \centering
  \includegraphics[width=\textwidth,height=1.9in,keepaspectratio]{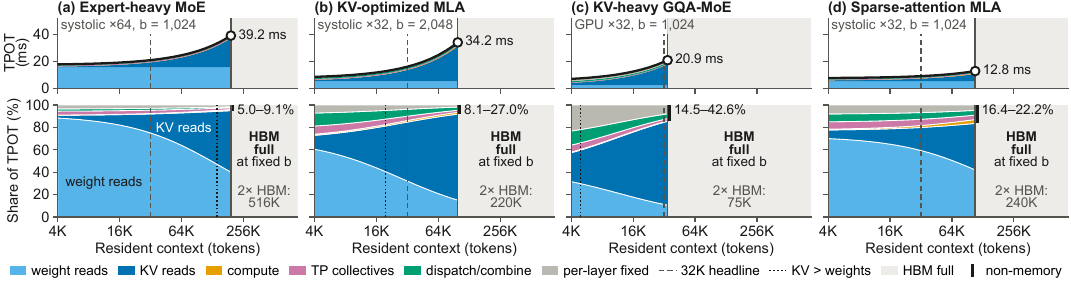}
  \vspace{-25pt}
  \caption{\textbf{Context shifts decoded-token time from weight reads toward
  KV work while absolute latency rises.} (a)--(d)~Each panel fixes
  one direction's selected P/D deployment (headline cell, electrical; decode
  class, units, and batch $b$ printed) and grows the resident context. Top:
  TPOT by component; bottom: exact additive shares (one microbatch). Gray zone:
  past the wall, $b$ exceeds the capacity-bound batch (label: wall at
  2$\times$ HBM capacity); re-batching serves longer contexts
  (Figure~\ref{fig:insights-rescue}). Bar: non-memory share at the wall
  (label: range).}
  \label{fig:insights-composition}
  \vspace{-10pt}
\end{figure*}

The extra class choices in \mbox{AFD$+$} change the
margin in only four of eight cases.
GPU placement helps P/D+AFD through the FFN pool in expert-heavy MoE and
through prefill in KV-heavy electrical. For example, expert-heavy MoE on
photonic fabric selects a GPU FFN pool under \mbox{AFD$+$}. Its margin improves by
$5.8$ points (\figpanel{fig:insights-2x2}{a}). P/D also benefits from
phase-split placement, which moves the KV-optimized photonic comparison
toward P/D (\figpanel{fig:insights-2x2}{b}). Extra class choices can improve
either design.

Pool allocation explains the positive headline margins. When optimized
alone under the same TPOT target, P/D has higher throughput
per decode kW in every cell. Selected best-allowed P/D+AFD deployments also
have longer TPOT than P/D, although both meet the target. KV-heavy GQA-MoE
on photonic fabric gives the largest P/D+AFD margin, $+8.1\%$
(\figpanel{fig:insights-2x2}{c}). Its pool
allocation leaves room for eleven prefill replicas, versus ten for
P/D. That replica explains the gain.

At the headline workload, sharing the attention/FFN link with in-pool
collectives (e.g., local TP reduction) does not limit the selected deployments. Dedicated and shared modes select the same configurations and
yield the same margins, giving eight distinct results from 16 comparisons.
At nominal values, the search selects systolic and GPU classes. Keeping all
FFN weights resident on the modeled SRAM/dataflow class exceeds the
128\,kW budget for these directions.

\begingroup\clubpenalty=10000

FastAFD measures 1.35--1.45$\times$ the per-GPU decode throughput of tuned
co-located vLLM~\cite{fu2026fastafd}. AFD-Ledger's model re-provisions both
designs under one budget and narrows AFD's Qwen3-235B 8K gain to
0.99$\times$~\cite{qiu2026afdledger}. On these development data, \tool{}
predicts 1.37$\times$ and 1.47$\times$ at FastAFD's recorded 8K and 16K
deployments. Optimizing each design within FastAFD's GPU budgets gives
0.94$\times$ at both. In these modeled cases, AFD's gain disappears
when both designs can optimize their mappings and pool sizes under the
same GPU budget.

\par\endgroup

Finer EP choices can avoid oversized decode pools, leaving more power
for prefill. The main power-of-two rule makes P/D resize its co-located
decode pool in larger steps, while P/D+AFD can size attention separately.
Allowing any integer EP for P/D (fractional experts per rank,
\S\ref{sec:model-equations}) removes most P/D+AFD regime-map leads beyond
9\% (\S\ref{sec:capacity}). This relaxes only P/D's sizing rule.
Giving both finer pool sizes restores some leads in the tested
regime-map cases, but no nominal headline P/D+AFD margin exceeds 9\%.
Starting EP below 8 has little effect at the headline workload.
With finer sizing for both designs, most surviving leads reflect lower
power at nearly the same absolute goodput, and a few improve it.

\textbf{Takeaway 1.} Headline margins lean P/D and depend on pool allocation,
search limits, and prefill assumptions. They establish
no robust deployment-level AFD advantage over P/D.

\subsection{Role-specific placement trades weight reads for exchange}
\label{sec:rolehw}\label{sec:role-hw}\label{sec:insights-demand}

Why does faster FFN hardware sometimes leave TPOT unchanged? Figure~\ref{fig:insights-composition} shows how TPOT changes with context for the four model directions, holding each direction's selected P/D deployment and batch size fixed. The upper strip shows TPOT by component in milliseconds, and the lower stack shows component shares. The gray region marks where that fixed batch no longer fits in memory.

\begin{figure*}[!t]
\vspace{-10pt}
\centering
\includegraphics[width=0.97\textwidth,height=1.90in,keepaspectratio]{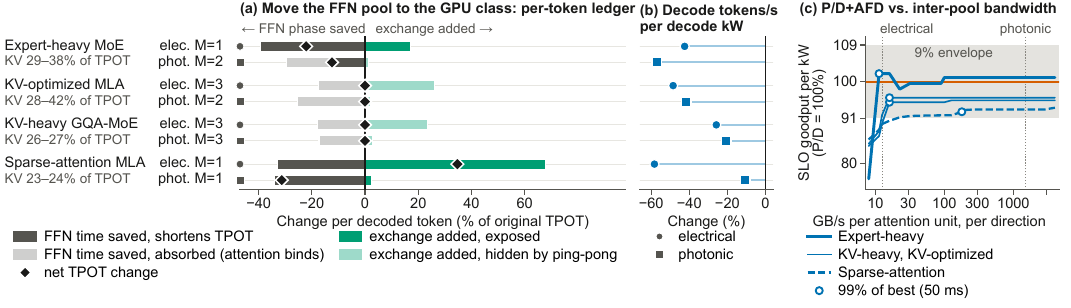}
\vspace{-11pt}
\caption{\textbf{Moving the FFN pool to GPUs saves FFN time equal to 17--39\% of original TPOT, yet every move loses decode tokens/s per decode kW.} Baselines in \figpanel{fig:insights-ledger}{a--b} are one-class P/D+AFD deployments selected under 128\,kW at 32K context, a 50\,ms TPOT target, and a dedicated link. (a)~Only FFN hardware changes. Row labels give this run's selected $M$, held fixed with batch size. Ledger in \% of original TPOT; KV-read time is unchanged. (b)~Change in decode tokens/s per decode kW; reassignment changes power. Circles mark electrical and squares photonic. (c)~Separate cross-class searches relative to P/D at a 50\,ms TPOT target (electrical, catalog domains, 4\,\textmu s). Circles mark 99\% of each direction's best goodput. Dotted lines mark the links in \figpanel{fig:insights-ledger}{a--b}.}
\label{fig:insights-ledger}\label{fig:link-budget}
\vspace{-12pt}
\end{figure*}

Weight-read time stays nearly constant per step, whereas KV work grows with context. Memory reads take 57--95\% of TPOT in these deployments. For example, KV reads overtake weight reads at about 5K tokens in the KV-heavy model direction (\figpanel{fig:insights-composition}{c}). Faster FFN weight reads then address a shrinking share of decode time. Sparse attention caps selected-token reads, but its index scan and full resident cache still grow (\figpanel{fig:insights-composition}{d}). All four fixed deployments run out of capacity before TPOT reaches the 50\,ms target. Longer contexts require re-batching or another deployment, evaluated in \S\ref{sec:capacity}.

\begingroup\setlength{\parskip}{1pt}
As \figpanel{fig:bg-timeline}{b} illustrates, faster FFN weight reads leave TPOT unchanged when attention remains the bottleneck. For this run, \figpanel{fig:insights-ledger}{a} starts from the one-class P/D+AFD deployment selected for each model--fabric case. The baseline search varies batch size and $M=1,\ldots,8$ along with hardware, mappings, and pool sizes. We then move only the FFN pool to the GPU class, keeping attention, prefill, the selected batch size and $M$, mappings, and pool sizes fixed. Pale bars show time absorbed by overlap. Dark bars show time that changes TPOT, and diamonds give the net change. For KV-optimized MLA on electrical fabric, three microbatches keep attention busy. The reduction in FFN service time equals 18\% of original TPOT. Added exchange equals 26\% of original TPOT, yet TPOT is unchanged. In this run, the selected sparse electrical deployment uses one microbatch. Without overlap, its TPOT grows by 35\% after the FFN move. In three other moves, TPOT falls. However, \figpanel{fig:insights-ledger}{b} shows that all eight moves reduce decode tokens/s per decode kW after accounting for the higher decode power of GPU reassignment. Shorter FFN service alone does not establish a more efficient deployment.

\begingroup\clubpenalty=10000
Inter-pool bandwidth helps until exchange no longer delays decode. For a homogeneous P/D+AFD stack on a dedicated link, hiding the exchange requires two conditions from Eq.~\eqref{eq:period}. A microbatch's full round trip must finish within the busier pool's work on all $M$ microbatches, and neither transfer may outlast that pool's work on one microbatch. \figpanel{fig:bg-timeline}{b} meets these conditions, but \figpanel{fig:bg-timeline}{c} does not. Unequal stages can leave enough slack with $M=2$, whereas balanced stages need at least three microbatches. At fixed microbatch size, extra microbatches keep more KV state resident. Fitting that state while still meeting the TPOT target can require more devices or smaller microbatches. With one microbatch, the selected sparse-attention deployments expose the round trip and leave both pools below 60\% utilization.

In \figpanel{fig:insights-ledger}{a}, every $M\geq2$ move already hides its exchange on its existing link. \figpanel{fig:insights-ledger}{c} shows what more bandwidth would buy. A separate search over cross-class P/D+AFD deployments sweeps the inter-pool bandwidth and reports SLO goodput per kW relative to P/D. Three of four model directions reach 99\% of their best goodput at 11--16\,GB/s per attention unit, per link direction (circles; 50\,ms TPOT target, catalog scale-up domains). Beyond this knee, extra bandwidth adds little goodput. Sparse attention gains only about 3\% more. At this 50\,ms target, no swept link produces a P/D+AFD lead beyond 9\%. At a tighter 10\,ms TPOT target, expert-heavy's nominal lead over its optimized P/D reference is 17.2\% at 8--200\,GB/s (not shown). That lead falls to 3.8\% with finer P/D pools. Hiding the exchange removes a cost but does not guarantee a deployment-level win~\cite{li2026opweave}.
\par\endgroup

In 8-unit scale-up domains (one 8-GPU server), the KV-heavy P/D decode pool spans several domains at 32K context and a 50\,ms TPOT target. Expert traffic across domains delays decode. P/D+AFD keeps its FFN pool in one domain. Its nominal lead is 9.2\%, falling to 2.8\% with finer P/D pools. No specification corners were tested. Larger domains or sufficiently fast links for both designs cut the lead below 9\%.

\begingroup\clubpenalty=10000
With the decode mappings of the 64 selected headline deployments (16 comparisons $\times$ two designs $\times$ two verdict sets) fixed, doubling memory bandwidth cuts TPOT by 39--49\%, versus at most 1.8\% for doubling FLOPs. Complete deployments can favor different upgrades because prefill also needs compute and pays for all-reduce and KV handoff. With uncached prompts, prefill uses 38--85\% of these deployments' power. Re-optimizing each design's hardware assignments, mappings, and pool sizes under 128\,kW can therefore make compute or link upgrades improve SLO goodput even when fixed decode barely responds.
\par\endgroup

\textbf{Takeaway 2.} FFN savings reduce TPOT when they shorten the limiting stage or round trip. Faster decode alone does not establish higher SLO goodput per kW, because prefill and power can change which upgrade helps.
\par\endgroup

\begin{figure*}[t]
  \vspace{-10pt}
  \centering
  \includegraphics[width=0.97\textwidth,height=1.85in,keepaspectratio]{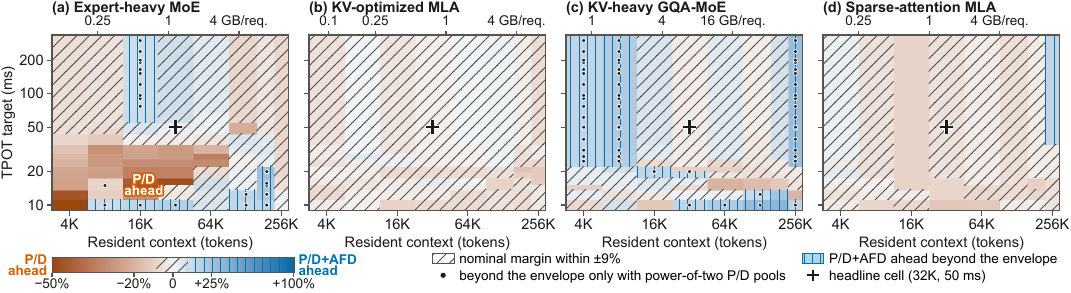}
  \vspace{-8pt}
  \caption{\textbf{Co-optimized at 128\,kW, P/D+AFD and P/D stay within the 9\% envelope in 579 of 768 cells; finer P/D pools preserve 18 of 101 P/D+AFD leads beyond it and 86 of 88 P/D leads.} Margin: best P/D+AFD over best P/D in SLO goodput per kW (nominal specifications; electrical, dedicated, TTFT 5\,s). Hatching: diagonal, within $\pm$9\%; vertical, P/D+AFD ahead beyond it. Top: KV per request. $+$:~headline cell (32K, 50\,ms); dots: beyond the envelope only with power-of-two P/D pools.}
  \label{fig:regime-map}
  \vspace{-12pt}
\end{figure*}

\begin{figure}[t]
  \centering
  \includegraphics[width=\columnwidth]{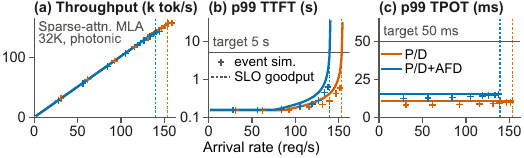}
  \vspace{-22pt}
  \caption{\textbf{At P/D's largest headline lead, P/D+AFD's 9.1\% lower absolute goodput
  comes from lower saturation throughput rather than different losses at the
  latency target. Both designs deliver 99\% of saturation throughput within the SLO.}}
  \label{fig:load-curves}
  \vspace{-12pt}
\end{figure}

%% file: sections/4b-Insights.tex
\subsection{Re-batching or reassignment restores feasibility}
\label{sec:context}\label{sec:capacity}\label{sec:insights-feasibility}

When context or arrival rate makes a fixed deployment infeasible, is the serving boundary at fault, or has one mapping exhausted its resources? We vary context, TPOT target, arrival rate, and KV footprint, then search offline for changes that restore feasibility. Figure~\ref{fig:regime-map} maps the margin for the four model directions over resident context (horizontal) and TPOT target (vertical). Diagonal hatching marks margins inside $\pm 9\%$, and vertical hatching marks P/D+AFD leads beyond it. The plus marks the 32K, 50\,ms headline cell. Dots mark leads beyond the envelope only with power-of-two P/D pools. P/D's leads beyond 9\% (unhatched) cluster at TPOT targets of 31\,ms or tighter and are largest in \figpanel{fig:regime-map}{a} and \figpanel{fig:regime-map}{c}. Prefill replicas and rounded pool sizes change in steps as context grows, shifting the margin. Context alone therefore does not give a single threshold for choosing the serving boundary.

\begin{figure}[!b]
  \vspace{-14pt}
  \centering
  \includegraphics[width=\columnwidth,height=2.21982645in]{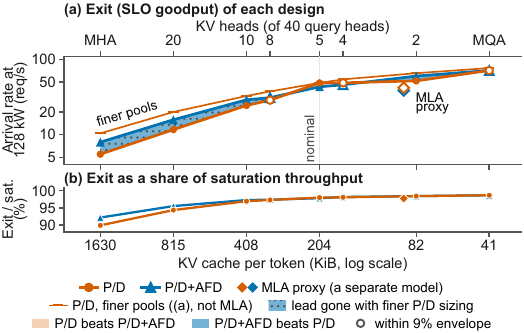}
  \vspace{-20pt}
  \caption{\textbf{Smaller KV caches raise both designs' SLO goodput; finer P/D pools erase the P/D+AFD leads in the MHA-to-MQA sweep.}
  (a)~Exits at TTFT 5\,s and TPOT 50\,ms, co-optimized on the GPU class under 128\,kW (KV-heavy GQA-MoE, 32K).
  Main curves use power-of-two EP; the thin line uses finer P/D pool sizing.
  (b)~Main-rule exit over saturation throughput.}
  \label{fig:insights-kv}
\end{figure}

We next separate throughput limits from latency-target failure. At P/D's largest headline lead, \figpanel{fig:load-curves}{a--c} plots throughput, p99 TTFT, and p99 TPOT against arrival rate. P/D+AFD's absolute goodput and saturation throughput are both 9.1\% lower; both designs deliver 99\% of saturation within the SLO. TTFT exceeds its target near saturation while TPOT stays below its target, so the goodput gap reflects lower saturation throughput. In Figure~\ref{fig:insights-kv}, we vary KV heads in the KV-heavy model direction, re-optimizing both designs on GPUs at 32K under 128\,kW. In \figpanel{fig:insights-kv}{a}, KV heads fall left to right from 40 (multi-head attention, MHA) to one (MQA)~\cite{shazeer2019mqa}. Each exit is SLO goodput; shading between the main curves marks rates where only the higher design meets both latency targets. Smaller caches cut per-request KV reads and storage, raising both load limits.

\begingroup\clubpenalty=10000
The thin P/D line allows finer decode-pool sizes, avoiding the main rule's coarse allocation steps. At MHA, P/D+AFD exits 45\% above main P/D but 24\% below P/D with finer pools. Only P/D receives this relaxation. The thin line exceeds P/D+AFD throughout the MHA-to-MQA sweep. \figpanel{fig:insights-kv}{b} shows both main configurations reaching 90--99\% of saturation. Their goodput gaps mainly reflect different saturation throughputs. The MLA diamonds represent a separate model that also changes attention computation. Unified serving, not drawn, reaches up to 2.6$\times$ the best split design's goodput. This is an upper bound because chunked prefill is fully hidden in decode steps (\S\ref{sec:inputs-axes}).
\par\endgroup

\begin{figure*}[t]
  \vspace{-10pt}
  \centering
  \includegraphics[width=\textwidth]{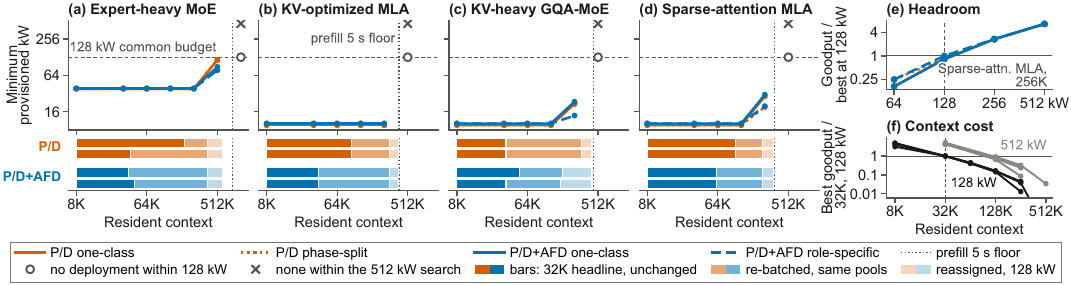}
  \vspace{-19pt}
  \caption{
\looseness=-1
\textbf{Re-batching or reassignment rescues every 32K headline deployment until both designs reach a TTFT wall at the same context.} (a)--(d)~Upper bars: best-allowed; reassignment stays within 128\,kW. Rescue does not restore short-context goodput. (e)~Budget: sparse-attention MLA at 256K, P/D+AFD's largest 128\,kW lead surviving finer P/D pools; 64-way prefill TP removes it. (f)~Best goodput at 128 and 512\,kW over the 32K best at 128\,kW, for four directions.}
  \label{fig:insights-rescue}
  \vspace{-10pt}
\end{figure*}

The rescue search tests whether re-batching, remapping, or hardware reassignment restores feasibility as context grows. Remapping changes mappings and pool sizes on the same classes, whereas reassignment can replace classes that miss TTFT. \figpanel{fig:insights-rescue}{a--d} plots resident context horizontally and the minimum provisioned power (kW) satisfying memory and SLO constraints vertically. Bars mark each 32K headline deployment's fixed, re-batched, and re-optimized limits. The last allows reassignment. Curves show the co-optimized power requirement. For example, the KV-heavy one-class P/D+AFD deployment (\figpanel{fig:insights-rescue}{c}) stops at 34K because its batch no longer fits in memory. Re-batching, a new batch and $M$ on the same pools, extends service to 168K. Full co-optimization extends both designs to 423K within 128\,kW. With uncached prompts, prefill eventually limits both designs at the same context within each direction. The KV-heavy example reaches this TTFT wall even though attention capacity holds to 1M. The 5\,s TTFT target becomes binding under our prefill rule (TP$\le$8 unless nothing smaller fits). Allowing prefill TP up to 64 makes 512K feasible for three directions.

Restoring feasibility does not restore 32K goodput. \figpanel{fig:insights-rescue}{e} sweeps the budget for sparse attention at 256K, where power raises both designs' goodput. \figpanel{fig:insights-rescue}{f} plots the best goodput at each context and budget as a fraction of the 32K best at 128\,kW. At 256K, increasing sparse attention's budget from 128 to 512\,kW raises that fraction from 1.3\% to 9\%. Per-request prefill time grows faster than context length.

The uncached-prompt wall does not apply unchanged to a cached prefix. We re-evaluate Wu et al.'s Qwen3-235B case with a cached 1M-token prefix (FP8 KV)~\cite{wu2026howfar}. Only the 4,096 new tokens are prefilled. Under the same capacity and performance rules for both designs, P/D fits on their B200. With our classes at 128\,kW, it also fits at nominal values and all 32 corners. Their ``AFD required'' conclusion therefore does not follow. Under the main EP rule at 128\,kW, best-allowed P/D+AFD leads at nominal values ($+32\%$) and at all 32 corners ($+5\%$ to $+99\%$). This lead is conditional (\S\ref{sec:cal-test}).

\textbf{Takeaway 3.} Context length alone does not choose the serving boundary. KV footprint, latency targets, and pool sizing shape both designs' supported load. Re-optimization can extend context limits, but uncached prefill can still leave long-context SLO goodput far below the short-context baseline.

\subsection{Hardware assignments are not stable}
\label{sec:insights-roles}\label{sec:regret}

Can a fleet commit each serving role to one accelerator class? Changing the objective from goodput per kW to goodput per deployment unit moves six of eight best deployments from systolic to GPU-based hardware. We measure lock-in by varying specification corner, context, model direction, or objective, changing only each nominal-best 128\,kW deployment's batch. With comparisons and axes weighted equally, the median loss is 35.7\% of scenario-best SLO goodput per kW (per unit on the objective axis). Re-optimizing mappings and pool sizes within the existing hardware cuts this loss to 9.3\%. Beyond remapping, a 25\% reserve of units or power lowers it to 5.1\%, and switching the serving design lowers it to 5.7\%. Doubling inter-pool bandwidth lowers it only to 9.2\%. None of these changes supplies an absent accelerator class.

\textbf{Takeaway 4.} The evidence supports capability preservation rather than early lock-in.
Fleet planners should let the serving decomposition and hardware assignment
follow the models they serve. They should keep mappings and pool sizes
adjustable rather than lock a future fleet to the accelerator class with the
highest score at one operating point.

%% file: sections/5-Calibration.tex
\section{Calibration and Robustness Testing}
\label{sec:calibration}

\subsection{Governing equations}
\label{sec:model}\label{sec:model-equations}\label{sec:tool-law}

Each role's service time follows its busiest chip's compute and memory work. Role $r\in\{\mathrm{A},\mathrm{F}\}$ has per-microbatch layer time $t_r=\max(\Phi_r,\Theta_r)+(1-\gamma)\min(\Phi_r,\Theta_r)+t_{\mathrm{TP}}+t_0$ in Eq.~\eqref{eq:period}. Here $\Phi_r$ and $\Theta_r$ are compute and memory times, $\gamma$ their calibrated overlap ratio, $t_{\mathrm{TP}}$ TP all-reduce time, and $t_0$ fixed per-layer launch latency. Each microbatch re-streams its weights and pays $t_0$.

\noindent\textbf{Exchange.}
Both designs pay for expert dispatch and combine across scale-up and scale-out tiers. Dispatch sends each token of width $d$ to $k_e=k+n_{\mathrm{sh}}$ experts (shared ones always active) at $\sigma_{\mathrm{D}}$ bytes per element. $t_{\mathrm{D}}$ adds per-hop latency for every touched scale-up domain to the slower of injection, $b_s k_e d\sigma_{\mathrm{D}}/\beta_{\mathrm{inj}}$, and ejection, $\iota\,b k_e d\sigma_{\mathrm{D}}/(\mathrm{EP}\,\beta_{\mathrm{ej}})$. Here $b_s$ is the sending rank's token count, $\beta_{\mathrm{inj}}$ and $\beta_{\mathrm{ej}}$ are endpoint bandwidths, and $\iota=1$ gives mean receiver load. Combine uses the same rule with $\sigma_{\mathrm{C}}=2\sigma_{\mathrm{D}}$ (FP8 dispatch, 16-bit combine). The model includes endpoint bandwidth limits without tracing switch paths or simulating network queues. Experts are spread evenly over EP ranks, fractionally when EP does not divide the expert count.

\noindent\textbf{Capacity.}
Memory available for KV state bounds concurrency. A unit of capacity $H$ has $\phi H-H_c$ usable memory, with $\phi=0.9$~\cite{vllm2026engineargs} and communication buffer $H_c=2$\,GiB; SRAM/dataflow uses its full SRAM. Resident requests $U$ are capped at $\mathrm{DP}_{\mathrm{A}}s_{\mathrm{kv}}(\phi H-H_c-\omega)/((C+O)\kappa)$. $\mathrm{DP}_{\mathrm{A}}$ counts attention replicas, $s_{\mathrm{kv}}$ is the KV shard factor, $\kappa$ KV bytes per token, $C+O$ the peak resident context after $O$ output tokens, and $\omega$ other resident state.

\noindent\textbf{Waiting and SLO goodput.}
We convert service times into SLO goodput with a queue per pool. For Poisson arrivals at rate $\lambda$ and $c$ servers of rate $\mu$, Erlang-C gives queueing probability $P_Q(c,\lambda/(c\mu))$~\cite{harcholbalter2013performance}. Conditional wait is exponential with Allen--Cunneen rate $2(c\mu-\lambda)/(1+v^2)$~\cite{allen1990probability}, where $v^2$ is the service-time squared coefficient of variation. Nominal $v^2=1$ is conservative for the fixed-length workload. Decode's $U$ slots of rate $1/(OT)$ saturate at $X/O$ requests/s. Prefill has $R_{\mathrm{P}}$ replicas of $b_{\mathrm{P}}$ slots at rate $1/t_{\mathrm{P}}$. Given $T\le S_{\mathrm{TPOT}}$ and $\lambda<c\mu$ in each pool, SLO goodput is the largest $\lambda$ satisfying $t_{\mathrm{P}}+t_{\mathrm{KV}}+T/M+W_{99}\le S_{\mathrm{TTFT}}$. $t_{\mathrm{KV}}$ is the KV handoff, $W_{99}$ the exact p99 of the summed prefill and decode waits $W_{\mathrm{P}}+W_{\mathrm{D}}$, and $T/M$ bounds the microbatch-boundary wait.

\noindent\textbf{Queue scope.}
The fixed-order schedule folds per-layer synchronization and attention--FFN backpressure into $T$. The queue omits batch coupling, stragglers, and decode-to-prefill backpressure, and treats the two waits as independent.

\subsection{Frozen calibration and validation}
\label{sec:validation}\label{sec:cal-fits}

\begin{figure}[t]
  \centering
  \includegraphics[width=\columnwidth]{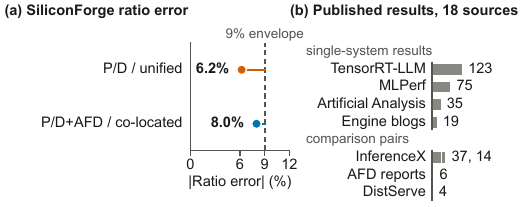}
  \vspace{-22pt}
  \caption{\textbf{Comparative errors against the 9\% envelope.}
  (a) Ratio errors (dots) and distance to the 9\% envelope (segments; \S\ref{sec:cal-test}).
  (b) Published prediction checks by source.}
  \label{fig:validation}
  \vspace{-12pt}
\end{figure}

We freeze four timing effects across all evaluation scenarios: GEMV-to-GEMM bandwidth Hill curves (per class where public data support them), a common sustained-to-peak compute ratio $\eta$, per-layer launch latency $t_0$, and compute/memory overlap $\gamma$. Calibration used published low-concurrency token rates without target-specific tuning. We compare the frozen model's throughput and between-design ratios~\cite{hill2019gables} with 252 throughput results and 47 primary comparison pairs from 18 published sources (\figpanel{fig:validation}{b}). The AFD and InferenceX~\cite{semianalysis2026inferencex} pairs were held out from tuning under a pre-registered protocol fixing the model and drivers before prediction. \figpanel{fig:validation}{a} compares errors with the 9\% envelope (\S\ref{sec:cal-test}).

The AFD checks use fixed configurations at saturated decode throughput without an SLO~\cite{fu2026fastafd,vllm2026afdplugin}, sometimes with forced expert balancing and logical-scale emulation. The primary check of unified and P/D frontiers at matched per-user rates~\cite{semianalysis2026inferencex} fails because under-predicted TTFT admits configurations excluded by the measured frontier. A later DistServe driver under-predicts P/D's advantage~\cite{zhong2024distserve}. Some published speedups were known before prediction. Absolute single-system throughput is substantially over-predicted~\cite{mlcommons2025inferv50,mlcommons2025inferv51,mlcommons2026inferv60,mlcommons2026inferv61,nvidia2026trtllmperf,artificialanalysis2025gpus,sglang2025largeep,sglang2025gb200p2,mooncake2025kimik2,vllm2026gb200,nvidia2025trtllmdsr1bestperf,nvidia2025trtllmblog1,nvidia2025trtllmgptoss,sglang2025gptoss}.

AFD checks cover fixed configurations within one scale-up domain; no published pair spans two studied accelerator classes. Other reports lack per-design steady-state goodput~\cite{amd2026moriio,wang2025taichi} or the realized mapping~\cite{weidman2026lpxlongctx,usami2026phase} and cannot be matched. Testing complete-deployment model estimates needs both designs' measured p99 TTFT, TPOT, steady-state goodput, and realized mapping near saturation on the same model, hardware, engine, and budget. An event-driven simulation checks only the queueing closed forms (\S\ref{sec:model-equations}). Goodput margins agree within 0.2 percentage points; closed forms conservatively estimate p99 TTFT near saturation.

\subsection{When a modeled difference is interpretable}
\label{sec:cal-test}\label{sec:decision-rule}

For headline comparisons, the \emph{margin} $\Delta$ in SLO goodput per kW is (P/D+AFD $-$ P/D)/P/D; positive values favor P/D+AFD. The two-part test calls a comparison \emph{decided}, with a \emph{robust} margin, when the same design leads at every tested specification corner and nominal $|\Delta|$ exceeds the \emph{9\% envelope}; otherwise, it is \emph{undecided}. Corners represent specification uncertainty, not a sampling distribution, and do not establish behavior between them. Published inference-error summaries anchor the 9\% comparison envelope~\cite{zhang2024llmcompass,agrawal2024vidur,feng2026frontier}. Verdicts passing 9\% but not the largest observed AFD gain error are \emph{conditional}, as are all role-specific verdicts because no published pair spans two studied classes.

\subsection{Applying the two-part test}
\label{sec:cal-demo}\label{sec:cal-corners}\label{sec:decide}\label{sec:decidability}

\begin{figure}[t]
  \centering
    \includegraphics[width=\columnwidth]{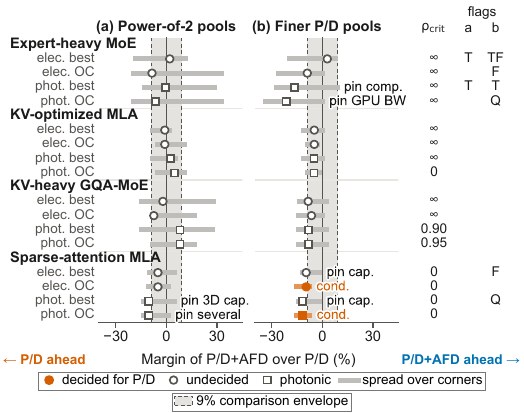}
  \vspace{-21pt}
  \caption{\textbf{Headline verdicts under both pool-sizing rules.} best: best-allowed; OC: one-class; cond.: conditional; pin: range deciding the row when pinned (BW: memory bandwidth; several: no single range); $\rho_\mathrm{crit}$ (\figpanel{fig:forest}{a}): least sign-preserving error correlation ($\infty$: none). Pre-declared \figpanel{fig:forest}{a--b} flags: T, within 10\% of the TPOT target; Q, verdict changes with $v^2{=}4$ or saturation throughput instead of SLO goodput; F, verdict changes with another collective or link convention.}
  \label{fig:forest}
  \vspace{-12pt}
\end{figure}

We independently re-optimize both designs at each tested specification corner. Figure~\ref{fig:forest} shows nominal margins (markers) and corner ranges (bars) against the 9\% envelope. Under the main rule (\figpanel{fig:forest}{a}), every range spans both signs; no headline comparison is decided. With finer P/D pools (\figpanel{fig:forest}{b}), one-class sparse attention conditionally favors P/D beyond 9\% on both fabrics. No P/D+AFD or best-allowed comparison is decided. Best-allowed sparse attention ties at one corner where the same prefill pool limits both designs.

\noindent\textbf{Error cancellation.}
We sweep error correlation in Monte Carlo with 9\% error per candidate deployment, re-selecting each design's best without extrapolating the six AFD pairs' measured cancellation. Figure~\ref{fig:forest}'s $\rho_\mathrm{crit}$ is the least correlation preserving the nominal lead in 95\% of draws ($\infty$: none).

\noindent\textbf{Robustness.}\label{sec:robust}
Under the main rule, halved $t_0$, $\gamma=0.5$ or $1.0$, $v^2$ up to 4, collective or link conventions, pipeline fill and drain, a straggler, capacity--bandwidth coupling, and $M\le2$ leave headline comparisons undecided but can move nominal margins. Less usable memory ($\phi=0.8$) can favor P/D+AFD, while prefix reuse can favor either design. Halved ranges decide photonic sparse attention for P/D. With finer P/D pools, speculative decoding (\S\ref{sec:discussion}) or full usable memory ($\phi=1$) can undo one-class preferences; all added verdicts favor P/D.

Whole-expert EP divisors give no nominal P/D+AFD lead beyond 9\%. Their main-rule subset caps KV-heavy's P/D pool at 64 units and can create such a lead.

%% file: sections/6-Discussion.tex
\section{Discussion and Scope}
\label{sec:discussion}
\label{sec:cal-scope}

\noindent\textbf{Serving and capacity model.}
Both designs use the same queueing and capacity model. Design-specific scheduler or prefix-reuse policies do not enter the margin. Other policies, parallelism, or distributed prefill may change feasibility.

\begingroup\clubpenalties 3 10000 10000 0
\noindent\textbf{Large expert parallelism can narrow what AFD adds.}
Published decode deployments use EP~72--320~\cite{sglang2025largeep,deepseek2025inferencesys,deepseek2024v3,zuo2025cloudmatrix}. Large EP gives co-located decode the FFN batching of AFD's second pool. At 512\,kW, forcing both designs to EP~64 or~128 can flip GPU one-class margins toward P/D. With EP re-optimized, electrical best-allowed margins remain within the 9\% envelope. We omit EP beyond the expert count and configurations with redundant experts.
\par\endgroup

\noindent\textbf{Expert imbalance is an evaluated sensitivity.}
\tool{} can estimate the busiest EP rank's load under uniform random routing with a binomial model, charging it to routed FFN work and receiving traffic. We evaluated this sensitivity with dedicated links at 32K context, at 128\,kW for one-class and best-allowed placement and at 512\,kW for GPU one-class placement. With EP re-optimized, no margin changes sign or its side of the 9\% envelope.

\noindent\textbf{New fabric features are link parameters.}
NVLink~5 and~6, SHARP, NVLink Fusion, and device-initiated RDMA~\cite{nvidia2026nvlinkpage,aubrey2026verarubin,nvidia2025nvlinkfusion,hamidouche2025gin} map to tier bandwidth, latency, domain size, and collective offload. SHARP speeds all-reduce, not dispatch and combine. In headline comparisons, no tested named link gives P/D+AFD a lead beyond 9\% under either pool-sizing rule. Small domains give AFD a nominal lead (\S\ref{sec:rolehw}).

\noindent\textbf{Extending to speculative and diffusion decoding.}
Speculative decoding can place a small draft and larger verifier in separate pools~\cite{gimlet2026specdecode} and increases both designs' FFN tokens per weight read~\cite{leviathan2023speculative,li2024eagle,deepseek2024v3}. In our one-draft-token sensitivity, which omits draft cost and draft or transient KV state, KV-optimized MLA's one-class AFD lead in throughput per decode kW begins at nearly twice the context. Full-serving SLO goodput includes prefill under the common budget; this sensitivity establishes no P/D+AFD preference under either pool-sizing rule (\S\ref{sec:cal-test}). Unevaluated diffusion decoding raises tokens per weight read~\cite{nie2025llada,inception2025mercury} and repeats prefill~\cite{kedia2026sangam,fan2025dllmserve}.

%% file: sections/8-RelatedWork.tex
\section{Related Work}\label{sec:related}

\noindent\textbf{Serving disaggregation.}
DistServe and Splitwise separate prefill from decode
\cite{zhong2024distserve,patel2024splitwise}; Mooncake emphasizes KV transfer
and reuse~\cite{qin2025mooncake}, and P/D-Serve and HexGen-2 demonstrate
large-scale and heterogeneous deployments
\cite{jin2024pdserve,jiang2025hexgen2}. Dynamo, SGLang, llm-d, and Corsair
provide production P/D paths~\cite{nvidia-dynamo,sglang2026pddocs,llmd2026,
dmatrix2026corsairproduction}. Heterogeneous-P/D
studies show accelerator class, precision, fabric, and KV residency change
the hardware assignment~\cite{wang2026heterogeneous,wei2026hmaserve}. These systems fix the boundary at P/D. Splitwise asks which silicon should serve each phase. We ask this at the attention/FFN boundary.

MegaScale-Infer, Step-3, xDeepServe, Janus, Lamina, AiDE, and Rubin--LPX use an attention/FFN split
\cite{zhu2025megascaleinfer,wang2025step3,xiao2025xdeepserve,zhang2025janus,chen2024lamina,kim2025aide,nvidia2026groq3lpx}.
Measurements, models, and sweeps show that AFD's value depends on the
model, workload, and fabric
\cite{wu2026howfar,song2026analytical,liu2026revealing,
mitra2025beyondbuzz,li2026ods}. AFD's measured gain changes sign with pool provisioning~\cite{vllm2026afdplugin}.
A gain at one deployment does not settle the boundary choice before future model demands and hardware resources are known.

\noindent\textbf{Closest design-space models.}
Wu et al.\ search parallelism, P/D, and AFD on 128 B200 GPUs, deferring unreleased hardware~\cite{wu2026howfar}; we re-evaluate one case in \S\ref{sec:capacity}.
Song et al.\ study A/F provisioning under stochastic workload variation using latency models calibrated on Huawei Ascend 910C NPUs~\cite{song2026analytical}. Their closed-form rule assumes a compute-bound FFN, with no co-located baseline. Their FFN arithmetic intensity is 299\,FLOP per weight byte. This lies below the FP8 ridge of B200 and our GPU and systolic classes, where the attention-to-FFN ratio scales with FFN memory bandwidth.
Frontier's simulations flip the P/D-versus-AFD ranking with the TTFT SLO~\cite{feng2026frontier}. Liu et al.\ trace an AFD ``dead zone'' to scale-out bandwidth~\cite{liu2026revealing}.
\looseness=-1
These four studies evaluate fixed hardware. Models and simulators cover current and prospective systems~\cite{williams2009roofline,hill2019gables,zhang2024llmcompass,bambhaniya2024genz,xu2026aiconfigurator,agrawal2024vidur,cho2026llmservingsim2,bambhaniya2026mist,chen2026fleetopt}. Statistical methods track model error and quantify predictive confidence over measured systems~\cite{chen2012statistical,yi2003statistically,wunderlich2003smarts,alameldeen2003variability,gutierrez2014sources,nowatzki2015simulators,balamurugan2026omnipilot,ray2025statistical}.
Prospective SPAD, Rubin CPX, memory-centric attention, and Raptor studies illustrate the need to combine these practices for hardware that cannot yet be run~\cite{zhang2025spad,nvidia2025rubincpx,park2024attacc,yu2026amma,nair2026raptor}. We vary future LLM architecture directions anchored in published families and hardware resource ranges anchored in announced accelerators (Tables~\ref{tab:models} and~\ref{tab:ledger}). Regime and pool-sizing searches recorded about 59.5 million decode candidate evaluations; both designs are re-optimized under a common budget at each specification corner. We compare nominal margins with model error and test whether their signs survive specification uncertainty.

\noindent\textbf{Concurrent work.}
Three studies posted after the initial submission report related conditions. OpWeave proves that one-class operator splitting pays only once KV growth pushes the co-located batch below the compute-bound threshold~\cite{li2026opweave}. HeteroPanacea finds that adding AFD to P/D pays only on genuinely different devices~\cite{forys2026heteropanacea}. Under a common budget, AFD-Ledger selects one-class AFD in 7 of 36 settings~\cite{qiu2026afdledger}. SQD instead splits decode at the quadratic/subquadratic attention boundary~\cite{tschand2026sqd}. Our comparison tests P/D against P/D+AFD across future LLM architecture directions and jointly varied hardware-specification ranges.

%% file: sections/9-Conclusion.tex
\section{Conclusion}\label{sec:conclusion}

\tool{} separates serving decomposition from hardware assignment. Headline comparisons lean P/D without a robust AFD advantage. AFD nominally leads for the KV-heavy direction when the co-located expert pool spans small scale-up domains.
Hardware assignments change with objectives and specification corners.
Prospective modeling informs substantial fleet investments that shape customer experience and accelerator roadmaps. We recommend keeping mappings
and pool sizes adjustable.

%% file: sections/13-AIUse.tex
\section*{AI use}\label{sec:ai-use}\label{sec:aiuse}
AI systems assisted with manuscript preparation and code development.
All AI-assisted text, code, and results were independently reviewed, revised, and verified by the authors, who take full responsibility for this paper.